\documentclass[journal]{IEEEtran}

\usepackage[acronym]{glossaries} 
\glsdisablehyper

\usepackage[hyphens]{url}
\usepackage[hidelinks]{hyperref}
\hypersetup{breaklinks=true}
\usepackage{cite}
\usepackage{comment}
\usepackage{mathtools}
\usepackage{amsmath,amsfonts,bm}
\usepackage{optidef}
\usepackage{algorithmic}
\usepackage{graphicx}
\usepackage{multirow}
\usepackage{multicol}
\usepackage{gensymb}
\usepackage{comment}
\usepackage{url}
\usepackage[utf8]{inputenc}

\DeclareMathOperator*{\argmax}{argmax} 

\newcommand{\bx}{{\bm x}}
\newcommand{\by}{{\bm y}}
\newcommand{\bz}{{\bm z}}

\newcommand{\bff}{{\bm f}}
\newcommand{\bmu}{{\bm \mu}}
\newcommand{\bSigma}{{\bm \Sigma}}
\newcommand{\beps}{{\bm \epsilon}}

\newcommand*\myglsentry[1]{%
  \protect\ifglsused{#1}{%
    \glsentryshort{#1}%
  }{%
    \glsentrylong{#1}%
  }%
}

\newacronym{ad}{AD}{Alzheimer's disease}
\newacronym{aac}{AAC}{augmentative and alternative communication}
\newacronym{als}{ALS}{amyotrophic lateral sclerosis}
\newacronym{asha}{ASHA}{American Speech-Language-Hearing Association}
\newacronym{asr}{ASR}{automatic speech recognition}
\newacronym{bci}{BCI}{brain computer interface}
\newacronym{cnn}{CNN}{convolutional neural network}
\newacronym{dnn}{DNN}{deep neural network}
\newacronym{dtw}{DTW}{dynamic time warping}
\newacronym{em}{EM}{expectation-maximisation}
\newacronym{dwm}{DWM}{digital waveguide mesh}
\newacronym{ecog}{ECoG}{electrocorticography}
\newacronym{eeg}{EEG}{electroencephalography}
\newacronym{ema}{EMA}{electromagnetic articulography}
\newacronym{emg}{EMG}{electromyography}
\newacronym{fem}{FEM}{finite element method}
\newacronym{fmri}{fMRI}{functional \myglsentry{mri}}
\newacronym{fnirs}{fNIRS}{functional near-infrared spectroscopy}
\newacronym{gmm}{GMM}{Gaussian mixture model}
\newacronym{gru}{GRU}{gated recurrent unit}
\newacronym{hmm}{HMM}{hidden Markov model}
\newacronym{hdr}{HDR}{haemodynamic response}
\newacronym{ifg}{IFG}{inferior frontal gyrus}
\newacronym{lfp}{LFP}{local field potential}
\newacronym{lis}{LIS}{lock-in syndrome}
\newacronym{lsf}{LSP}{line spectral pair}
\newacronym{lstm}{LSTM}{long short-term memory}
\newacronym{lvcsr}{LVCSR}{large vocabulary continuous speech recognition}
\newacronym{meg}{MEG}{magnetoencephalography}
\newacronym{mfcc}{MFCC}{Mel-frequency cepstral coefficient}
\newacronym{mlpg}{MLPG}{maximum-likelihood parameter generation}
\newacronym{mle}{MLE}{maximum likelihood estimation}
\newacronym{mllr}{MLLR}{maximum likelihood linear regression}
\newacronym{mse}{MSE}{mean squared error}
\newacronym{mmse}{MMSE}{minimum \myglsentry{mse}}
\newacronym{mri}{MRI}{magnetic resonance imaging}
\newacronym{nidcd}{NIDCD}{National Institute on Deafness and Other Communication Disorders}
\newacronym{nnlm}{NNLM}{neural network language modelling}
\newacronym{pca}{PCA}{principal component analysis}
\newacronym{pd}{PD}{Parkinson's disease}
\newacronym{pma}{PMA}{permanent magnetic articulography}
\newacronym{rnn}{RNN}{recurrent neural network}
\newacronym{rtmri}{rtMRI}{real-time \myglsentry{mri}}
\newacronym{semg}{sEMG}{surface \myglsentry{emg}}
\newacronym{slt}{SLT}{speech-language therapy}
\newacronym{ssi}{SSI}{silent speech interface}
\newacronym{stg}{STG}{superior temporal gyrus}
\newacronym{tts}{TTS}{text-to-speech}
\newacronym{voca}{VOCA}{voice-output communication aid}
\newacronym{vsmc}{vSMC}{ventral sensorimotor cortex}
\newacronym{who}{WHO}{World Health Organization}
\newacronym{wfst}{WFST}{weighted finite-state transducers}
\newacronym{wer}{WER}{word error rate}

\usepackage{textcomp}
\def\BibTeX{{\rm B\kern-.05em{\sc i\kern-.025em b}\kern-.08em
    T\kern-.1667em\lower.7ex\hbox{E}\kern-.125emX}}
   
\begin{document}

\title{Silent Speech Interfaces for Speech Restoration: A Review}
\author{Jose A. Gonzalez-Lopez,
		Alejandro Gomez-Alanis,
		Juan M. Martín-Doñas,
		José L. Pérez-Córdoba,
		and Angel M. Gomez%
 
\thanks{These authors are with the Department of Signal Processing, Telematics and Communications, University of Granada, 18071 Granada, Spain (email: \{joseangl,agomezalanis,mdjuamart,jlpc,amgg\}@ugr.es).}
}

\maketitle


\begin{abstract}
This review summarises the status of silent speech interface (SSI) research. SSIs rely on non-acoustic biosignals generated by the human body during speech production to enable communication whenever normal verbal communication is not possible or not desirable. In this review, we focus on the first case and present latest SSI research aimed at providing new alternative and augmentative communication methods for persons with severe speech disorders. SSIs can employ a variety of biosignals to enable silent communication, such as electrophysiological recordings of neural activity, electromyographic (EMG) recordings of vocal tract movements or the direct tracking of articulator movements using imaging techniques. Depending on the disorder, some sensing techniques may be better suited than others to capture speech-related information. For instance, EMG and imaging techniques are well suited for laryngectomised patients, whose vocal tract remains almost intact but are unable to speak after the removal of the vocal folds, but fail for severely paralysed individuals. From the biosignals, SSIs decode the intended message, using automatic speech recognition or speech synthesis algorithms. Despite considerable advances in recent years, most present-day SSIs have only been validated in laboratory settings for healthy users. Thus, as discussed in this paper, a number of challenges remain to be addressed in future research before SSIs can be promoted to real-world applications. If these issues can be addressed successfully, future SSIs will improve the lives of persons with severe speech impairments by restoring their communication capabilities.
\end{abstract}


\section{Introduction}
\label{sec:intro}
\IEEEPARstart{S}{peech} is the most convenient and natural form of human communication. Unfortunately, normal speech communication is not always possible. For example, persons who suffer traumatic injuries, laryngeal cancer or neurodegenerative disorders may lose the ability to speak. The prevalence of this type of disability is significant, as evidenced by several studies. For instance, in \cite{Dupre03}, the authors conclude that approximately 0.4\% of the European population have a speech impediment, while a survey conducted in 2011 \cite{Eurostat11} concluded that 0.5\% of persons in Europe presented `difficulties' with communication. The \gls{asha} reports that nearly 40 million U.S. citizens have communication disorders, costing the U.S. approximately \$154-186 billion annually \cite{ASHA}. The \gls{who}, in its World Report on Disability \cite{WHO11} derived from a survey conducted in 70 countries, concluded that 3.6\% of the population had severe to extreme difficulty with participation in the community, a condition which includes communication impairment as a specific case. 

Speech and language impairments have a profound impact on the lives of people who suffer them, leading them to struggle with daily communication routines. Besides, many service and health-care providers are not trained to interact with speech-disabled persons, and feel uncomfortable or ineffective in communicating with them, which aggravates the stigmatisation of this population \cite{WHO11}. As a result, people with speech impairments often develop feelings of personal isolation and social withdrawal, which can lead to clinical depression \cite{Smith1996,Millard2001,Hunt2006,Yaruss2010,Byrne93, Braz2005, Danker2010}. Furthermore, some of these persons also develop feelings of loss of identity after losing their voice \cite{Shadden2005}. Communication impairment can also have important economic consequences if they lead to occupational disability.

In the absence of clinical procedures for repairing the damage originating speech impediments, various methods can be used to restore communication. One such is assistive technology. The U.S. \gls{nidcd} defines this as any device that helps a person with hearing loss or a voice, speech or language disorder to communicate \cite{NIDCD}. For the specific case of communication disorders, devices used to supplement or replace speech are known as \gls{aac} devices. \Gls{aac} devices are diverse and can range from simple paper and pencil resources to picture boards or \gls{tts} software. From an economic standpoint, the worldwide market for \gls{aac} devices is expected to grow at an annual rate of 8.0\% during the next five years, from \$225.8 million in 2019 to \$307.7 million in 2025 \cite{MarketWatchAAC}. \Gls{aac} users include individuals with a variety of conditions, whether congenital (e.g., cerebral palsy, intellectual disability) or acquired (e.g., laryngectomy, neurodegenerative disease or traumatic brain injury) \cite{Huer90, Beukelman07}.

In recent years, a promising new \gls{aac} approach has emerged: \glspl{ssi} \cite{Denby2010,Schultz2017,Denby2017}. \glspl{ssi} are assistive devices to restore oral communication by decoding speech from non-acoustic biosignals generated during speech production. A well-known form of silent speech communication is lip reading. A variety of sensing modalities have been investigated to capture speech-related biosignals, such as vocal tract imaging \cite{Hueber2010,Wand2016,Chung17}, electromagnetic articulography (magnetic tracing of the speech articulator movements) \cite{Schonle87,Fagan2008,Gonzalez2016,Cheah2016,Bocquelet2016}, \gls{semg} \cite{Schultz2010,Wand2011,Wand2014,Janke2017}, which captures electrical activity driving the facial muscles using surface electrodes, and \gls{eeg} \cite{Guenther2009,Herff2015,Anumanchipalli2019}, which captures neural activity in anatomical regions of the brain involved in speech production. The latter approach, involving the use of brain activity recordings, is also known as a \gls{bci} \cite{Wolpaw2002,Lebedev2006,NicolasAlonso2012}. Since \glspl{ssi} enable speech communication without relying on the acoustic signal, they offer a fundamentally new means of restoring communication capabilities to persons with speech impairments. Apart from clinical uses, other potential applications of this technology include providing privacy, enabling telephone conversations to be held without being overheard by bystanders and enhancing normal spoken communication in noisy environments \cite{Denby2010,Krishna2020}. These applications are possible because biosignals are largely insensitive to environmental noise and are independent of the acoustic speech signal (i.e., these biosignals can be captured even when no vocalisation is performed).

\glspl{ssi} have attracted increasing attention in recent years, as evidenced by the special sessions organised on this topic at related conferences \cite{SpecialSession09,SpecialSession15,SpecialSession18} and by special issues of journals \cite{Denby2010,SpecialIssue17}. These events and publications supplement the existing literature in the related research field of \glspl{bci} \cite{Wolpaw2002,Lebedev2006,Birbaumer2007,Brumberg2010,Lebedev2014,Chakrabarti2015,Herff2016,Brumberg2018}. In this review, we present an overview of recent advances in the rapidly evolving field of \glspl{ssi} with special emphasis on a particular clinical application: communication restoration for speech-disabled individuals. The remainder of this paper is structured as follows. Section \ref{sec:disorders} first summarises the speech and voice disorders that may affect spoken human communication, describing their causes and effects, and examines methods currently used to supplement and/or restore communication. Section \ref{sec:ssi} then formally introduces \glspl{ssi} and details the two main approaches employed in decoding speech from biosignals. The sensing modalities that have been proposed for capturing biosignals are described in Section \ref{sec:sensing_techniques}, which also provides an overview of previous research studies in which these sensing technologies have been used. Section \ref{sec:challenges} discusses the current challenges of \gls{ssi} technology and areas for future improvement. Finally, Section \ref{sec:conclusions} presents the main conclusions drawn from our analysis.


\section{Speech and language disorders}
\label{sec:disorders}
Human vocal communication is an extremely complex process involving multiple organs, including the tongue, lips, jaw, vocal cords and lungs, and requires precise coordination between these organs to produce specific sounds conveying meaning (phones). Vocal communication, however, can become difficult or even impossible when these organs, the anatomical areas in the brain involved in speech production or the neural pathways by which the brain controls the muscles are damaged or altered.

Table \ref{tab:disorders} presents a summary of the main types of disorders affecting spoken communication, their major causes and symptoms. As will be discussed later in more detail, a \gls{ssi} requires the acquisition of biosignals generated by the human body from which speech can be decoded. Depending on the specific disorder, some of the sensor technologies described in Section \ref{sec:sensing_techniques} may be better suited than others to capture such biosignals. For instance, sensors aimed at capturing the electrical activity in the language-related areas of the brain or the electrical activity driving the facial muscles will be better suited for people with dysarthria, who have difficulties moving and coordinating the lips and tongue, than using sensors for articulator motion capture (e.g., video cameras). The information about the applicable sensor technologies for each disorder is also shown in Table \ref{tab:disorders}.

\begin{table*}[t!]
\caption{Summary of the main disorders affecting spoken communication, their causes, effects and applicable sensor technologies to capture biosignals during speech production for individuals with these disorders.}
\centering
\begin{tabular}{llll}
\textbf{Disorder} & \
\textbf{Causes} & \textbf{Symptoms}               & \textbf{Applicable sensor technologies}\\ 
\hline
\hline
Aphasia & Brain injury caused by: & Difficulties with:  & Brain activity sensors \\
& - Stroke          & - Understanding language          &  \\
& - Head trauma     & - Speaking                        & \\
& - Brain tumors    & - Reading                         & \\
& - Infections      & - Writing                         &\\
& - Neurodegenerative diseases & &\\
\hline
Apraxia & Brain injury & Difficulties with: & Brain activity sensors\\
&                                              & - Moving and coordinating the articulators &  \\
&                                              & - Speak more slowly & \\
&                                              & - Unable to speak (severe cases) & \\
\hline
Dysarthria & Brain injury & Difficulty with lip \& tongue movements                    & Brain activity sensors\\
           &              & Speech is: & Muscle activity sensors\\
           &              & - Slurred, mumbled or choppy & \\
           &              & - Hard to understand & \\
           &              & - Monotone & \\
           &              & - Very loud or quiet & \\
\hline
Laryngectomy & Laryngeal cancer   & Voice loss    & Brain activity sensors\\
             & Severe neck injury &                & Muscle activity sensors\\
             &                      &                &  Articulator motion capture  \\
\hline
\end{tabular}
\label{tab:disorders}
\end{table*}

In the rest of this section, we provide an overview of the different types of speech, language and voice disorders, discuss their causes and describe methods and devices currently available to help speech-impaired people communicate, including previous investigations in which \glspl{ssi} have been used to restore communication.

\subsection{Aphasia}
\label{ssec:lang_disorders}
Aphasia is a disorder that affects the comprehension and formulation of language and is caused by damage to the areas of the brain involved in language \cite{Damasio92}. People with aphasia have difficulties with understanding, speaking, reading or writing, but their intelligence is normally unaffected. For instance, aphasic patients struggle in retrieving the words they want to say, a condition known as \emph{anomia}. The opposite mental process, i.e., the transformation of messages heard or read into an internal message, is also affected in aphasia. Aphasia affects not only spoken but also written communication (reading/writing) and visual language (e.g., sign languages) \cite{Damasio92}.

The major causes of aphasia are stroke, head injury, cerebral tumours or neurodegenerative disorders such as \gls{ad} \cite{Gorno2011}. Among these causes, strokes alone account for most new cases of aphasia \cite{NAA}. Elderly people are especially liable to develop aphasia because the risk of stroke increases with age \cite{Truelsen2006}. Aphasia affects the sexes almost equally, although the incidence is slightly higher in women \cite{Wallentin2018}.

The recommended treatment for aphasia is \gls{slt} \cite{NHSAphasia}. This includes restorative therapy, aimed at improving or restoring impaired communication capabilities, and compensatory therapy, based on the use of alternative strategies (such as body language) or communication aids to compensate for lost communication capabilities. These communication aids range from simple communication boards, where the patient points at the word, letter or pictogram required, to speech-generating electronic devices known as \glspl{voca} \cite{Beukelman2013}. These devices, however, are limited by their slow communication rates and are inappropriate for deeply paralysed patients.

In recent years, \glspl{bci} have gained considerable attention as a promising and radically new approach to restore communication to aphasic patients. Thus, in \cite{Kleih2016} a pilot study was conducted in which five persons diagnosed with post-stroke aphasia used a visual speller system for communication. The system consisted of a 6 $\times$ 6 matrix shown on a screen containing the alphabet letters and the digits 0 to 9. The matrix columns and rows were flashed randomly and an intended target cell was selected if P300 evoked potentials \cite{Farwell1988} were generated after the corresponding row and column were flashed. All participants were able to use the system with up to 100\% accuracy when spelling individual words. In \cite{Shih2013}, aphasic patients and healthy controls were compared using a P300 visual speller. Although the controls achieved significantly higher spelling accuracy than the aphasic subjects, these patients were able to use the system successfully to communicate. These \gls{bci} devices, however, are slow, cognitively demanding and unintuitive: communication rates up to 10-12 words/min are reported in \cite{Wolpaw2002} and mastering the control of these \glspl{bci} is an arduous task that can take several months of practice. In contrast, \glspl{ssi} have emerged as a more natural and promising alternative to restore oral communication to aphasic patients. Despite the literature on this topic is scarce, in \cite{Bocquelet2016b} the authors discussed the main considerations that should be taken into account when designing a \gls{ssi}-based systems for speech rehabilitation in aphasia. The interested reader can find an in-depth review of this topic in \cite{Chaudhary2016}.

\subsection{Apraxia and dysarthria}
\label{ssec:speech_disorders}
Apraxia and dysarthria are motor speech disorders \cite{Duffy2013} which are characterised by difficulties in speech production. To speak, the brain needs to plan the sequence of muscle movements that will result in the desired speech signal and to coordinate these movements by sending messages through the nerves to the relevant muscles. Unfortunately, this process is impaired in some individuals. For instance, apraxia of speech is caused by damage in the motor areas of the brain responsible for planning or programming the articulator movements \cite{Duffy2013}. Unlike aphasia, language skills are not impaired in persons with apraxia, although it often coexists with aphasia and other speech and language disorders. Dysarthria, on the other hand, is a motor speech disorder characterised by poor control over the muscles due to central or peripheral nervous system abnormalities. This often results in muscle paralysis, weakness or incoordination \cite{Darley1969}. Acoustically, dysarthric speech sounds monotonous, slow and, more importantly, is significantly less intelligible than normal speech \cite{Kent1999}.

Motor speech disorders account for a significant portion of the communication disorders in \gls{slt} practice. In fact, dysarthria is the most commonly acquired speech disorder, with an incidence of 170 cases per 100,000 population \cite{Enderby1996,Christensen2012}. Dysarthria occurs in about 25\% of post-stroke patients and 33\% of patients with traumatic brain injury \cite{Duffy2013}. The speech of patients with \gls{pd} is also affected by speech disorders, with about 60\% of them developing dysarthria \cite{Longemann78,Duffy2013}. This condition is also associated with other neurodegenerative disorders, such as \gls{als}, affecting more than 80\% of these patients \cite{Tomik2010}. In the worst case, \gls{als} patients with locked-in syndrome \cite{Smith2005}, a condition in which they are fully aware but suffer the complete paralysis of nearly all voluntary muscles, are practically unable to move or communicate. Apraxia is also common among patients with neurological conditions. \cite{Donkervoort2000} reported that one third of left hemisphere stroke patients have apraxia. This is also associated with dementia, occurring in about 35\% of patients with mild \gls{ad}, 50\% of those with moderate \gls{ad} and 98\% of those who are severely affected \cite{Edwards1991}. 

Individuals with these speech disorders may need to use \gls{aac} aids to compensate for communication deficits. Nevertheless, \gls{aac} aids requiring physical control are not always a viable solution because these disorders often co-occur with a physical disability. This makes \glspl{ssi} an attractive alternative to traditional \gls{aac} technology. Thus, recently, several studies have investigated the use of \glspl{ssi} for persons with motor speech disorders. In \cite{Deng2009dis}, \gls{asr} from \gls{emg} signals was investigated for dysarthric speakers. Parallel data with audio and \gls{emg} signals were recorded for dysarthric speakers while individual words were uttered. Then, \gls{asr} systems (see Section \ref{ssec:asr} for an overview of this \gls{ssi} approach) were independently trained for each modality. Experimental results show that speaker-independent \gls{asr} systems perform this task very poorly, whereas speaker-dependent acoustic models tailored to each subject yield the best scores, obtaining average word recognition accuracy of 95\% for audio-only, 85\% for \gls{emg} data, and 96\% for the combined modalities. One limitation of speaker-dependent \gls{asr} systems is that they require large amounts of data for training, which is not normally available in practice. To address this issue, \cite{Hahm2015} investigated articulatory normalisation, seeking to reduce the degree of variation in articulatory patterns across different speakers as a prior step to training speaker-independent \gls{asr} models. This technique was found to be highly effective, especially when acoustic and articulatory data were combined for speech recognition. To sum up, studies have demonstrated the viability of performing silent speech recognition on articulatory data captured from dysarthric speakers.

\subsection{Voice disorders}
\label{ssec:voice_disorders}
The vocal folds, or vocal cords, are two folds of tissue inside the larynx that play a major role in speech production. The rhythmic vibration of the vocal folds produced by airflow from the lungs is what creates the sound source (voice) in speech. This sound source is later shaped by the mouth and the nasal cavity to create the different phones in a language. 

Voice disorders are experienced when there is a disturbance in the vocal folds or any other organ involved in voice production, due to excessive or improper use of the voice, from trauma to the larynx or from neurological conditions such as \gls{pd} \cite{ASHAVoiceDisorder}. Perceptually, voice disorders are characterised by a hoarse voice (dysphonia) with altered pitch, loudness or vocal quality. In severe cases, e.g., patients who have their entire larynx removed to treat throat cancer (laryngectomy), voice disorders can even provoke the complete loss of voice \cite{Singer1981,Hilgers1990}. The reason for this is that the windpipe (trachea) is no longer connected to the mouth and nose after the laryngectomy, and so these patients can no longer produce the required sound to speak. Given the relatively high incidence of laryngeal cancer (it accounts for 3\% of all cancers \cite{CancerResearchUKCancer} with around 60\% of laryngectomised patients surviving five years or more \cite{Papadas2010}) and its devastating consequences, we focus on this disorder in the rest of the section.


Currently, there are three main options for speaking after a total laryngectomy: oesophageal speech, voice prosthesis and the electrolarynx \cite{CancerResearchUKSpeaking}. These three methods, however, are not without limitations \cite{Rameau2020}. In general, substitute voices generated by these methods are not agreeable, cannot be adequately modulated in terms of pitch and volume and are difficult to understand \cite{Maddalena1991,Miralles1995}. Moreover, women often dislike their new voice, finding it masculine and disturbing, due to the hoarse, deep nature of the sounds produced \cite{Sluis2019}. Furthermore, tracheoesophageal speech requires frequent valve replacement (every 3-4 months) as the valve may fail after becoming colonised by biofilm \cite{Ell96}. Oesophageal speech, on the other hand, is a skill that is difficult to master, requiring intensive, prolonged \gls{slt} and only about a third of the patients are able to master it \cite{Prakasan2019}. Finally, the voice generated by an electrolarynx sounds robotic and requires the patient to hold an external device, pressing it against the neck \cite{Fagan2008}. The drawbacks associated with each of these techniques negatively affect the patient's quality of life in terms of imperfect voice acceptance, restricted communication and limited social interaction \cite{Carr2000,Sluis2019}.

In contrast, \gls{ssi}-based speech restoration promises to overcome many of the above issues. Thus, the possibility of recognising speech from speech-related biosignals has been demonstrated for a variety of sensing techniques, such as \gls{semg} \cite{MaierHein2005,Jou2006,Schultz2010,Wand2014,MWand2018,Rameau2020}, \gls{ema} \cite{Wrench2000,Frankel2001}, \gls{pma} \cite{Fagan2008,Gilbert2010,Hofe2013,Cheah2016,Kim2017}, and imaging technologies based on video and ultrasound \cite{Hueber2008,Hueber2010,Tatulli2017}. Furthermore, direct speech generation from the captured biosignals (see Section \ref{ssec:ds}) is another possibility, having this approach the potential to restore the person's own voice, if enough recordings of the pre-laryngectomy voice are available for training \cite{Gilbert2017}. This second approach has been also validated for various modalities, including \gls{semg} \cite{AToth2009,Zahner2014,LDiener2015,Janke2017,diener2018comparison}, \gls{pma} \cite{Gonzalez2016,Gilbert2017,Gonzalez2017,Gonzalez2017b,Cao2018}, video-and-ultrasound \cite{Hueber2011,Hueber2012,Hueber2016} and Doppler signals \cite{Toth2010}. To sum up, the foundations have been laid for a future \gls{ssi}-based device for post-laryngectomy speech rehabilitation. However, with some notable exceptions \cite{Meltzner2017sil,Rameau2020}, the above proposals have been validated only for healthy users. In Section \ref{sec:challenges}, we discuss these and other challenges that need to be addressed in the near future.

\section{Silent speech interfaces}
\label{sec:ssi}
\begin{figure*}[ht!]
    \centering
    \includegraphics[width=\linewidth]{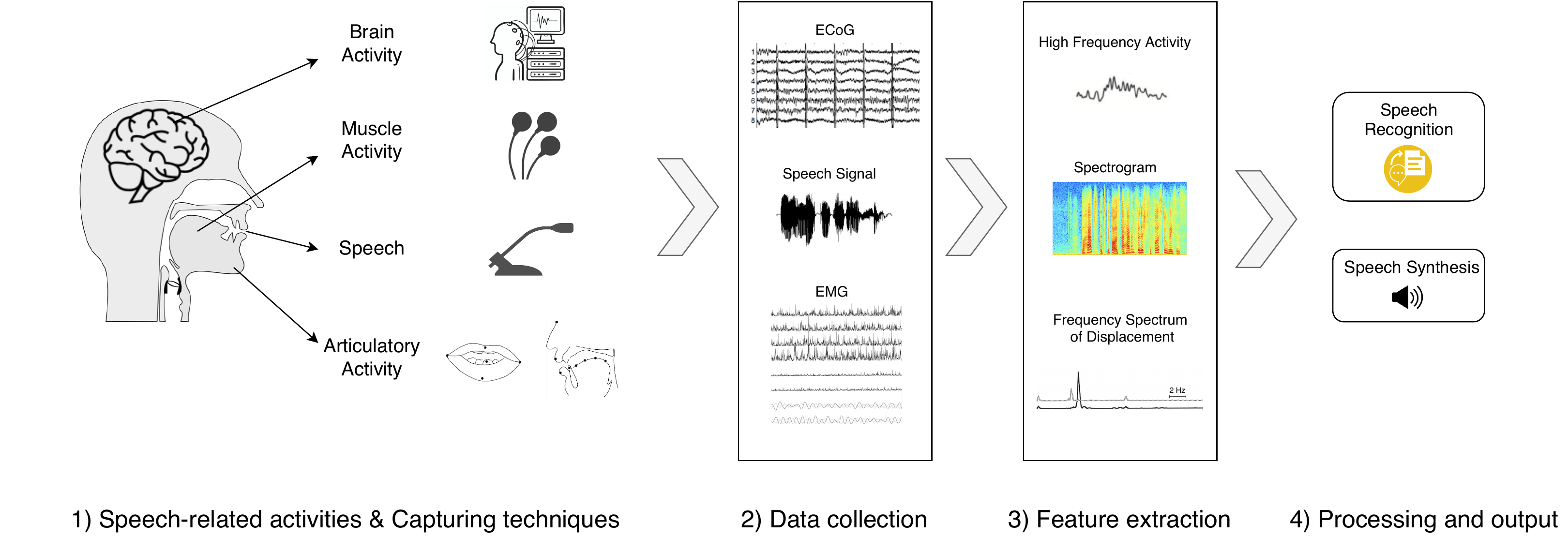}
    \caption{Block diagram of a \gls{ssi}-based communication system. First, speech-related activities during speech production are monitored with specialised sensing techniques (1). These activities produce a range of speech-related biosignals (2). Signal processing techniques are then used to extract a set of meaningful features from the biosignals (3). Finally, speech is decoded from these features, by \gls{asr} or direct speech synthesis (4). Figure adapted from \cite{Schultz2017}.}
    \label{fig:biosignals} 
\end{figure*}

As briefly introduced above, \glspl{ssi} are a new type of assistive technology for restoring oral communication. These devices exploit the fact that, in addition to the acoustic signal, other biosignals are generated during speech production, by different organs. These biosignals are the product of chemical, electrical, physical and biological processes that take place during speech production. As illustrated in Fig. \ref{fig:biosignals}, these processes include neural activity in the anatomical regions of the brain involved in speech planning and articulator motor control, activity in the peripheral nervous system providing motor control to the articulator muscles, articulatory gestures such as mouth opening or tongue movements, the vibration of the vocal folds (phonation) and pulmonary activity of the lungs during breathing. Depending on the specific disorder affecting the person, some of these processes might be disrupted whereas others will continue as usual. Consequently, the biosignals stemming from the unimpaired processes can be captured by sensing technologies, as detailed in Section \ref{sec:sensing_techniques}. 

As shown in Fig. \ref{fig:biosignals}, regardless of the type of biosignal considered, two \gls{ssi} approaches may be used to decode speech from this source \cite{Schultz2017,Gonzalez2017}: silent speech-to-text and direct speech synthesis. In the first approach, speech recognition algorithms \cite{Rabiner93,Huang2001,Yu2014} trained on silent speech data are used to decode speech from the feature vectors extracted from the biosignals. \Gls{tts} software \cite{Taylor2009,Zen2009} can then be used to synthesise speech from the decoded text if required. In the second approach, audible speech is directly generated from the biosignal features without an intermediate recognition step. Most commonly, \glspl{dnn} \cite{Lecun2015,Schmidhuber2015} trained on time-aligned speech and biosignal recordings (i.e., parallel data) are used to model the mapping between the biosignals and the acoustic speech parameters. In the following sections, these two \gls{ssi} approaches are described in greater detail.

\subsection{Silent speech to text}
\label{ssec:asr}
The goal of this \gls{ssi} approach is to convert speech-related biosignals into text (i.e., into a sequence of written words $\mathbf{w}= (w_1, \ldots, w_K)$). Normally, as illustrated in Fig. \ref{fig:biosignals}, \gls{asr} is not performed on the raw biosignals directly; instead, a more compact and parsimonious representation known as feature vectors is used. Thus, the raw biosignals are first pre-processed and converted into a sequence of feature vectors $\mathbf{X} = (\bx_1^\top, \ldots, \bx_T^\top)^\top$, where $\bx_t$ represents the $t$-th feature vector of dimension $D_x$ (i.e., column vector), and $T$ is the number of frames into which the input biosignal is divided. The computation of $\mathbf{X}$ depends on the type of biosignal. For instance, \glspl{mfcc} have been widely used for standard audio-based \gls{asr}, but other feature types need to be extracted for different biosignals, such as image-specific features for lipreading \cite{Chiou1997} or 3D coordinates of the speech articulators for \gls{ema} and \gls{pma} systems \cite{Wrench2000,Gilbert2010}. More details about specific biosignal features are given in Section \ref{sec:sensing_techniques}.

To determine the most likely sequence of words $\mathbf{\hat{w}}$ from $\mathbf{X}$, the following optimisation problem must be solved
\begin{equation}
\label{eq:asr}
    \mathbf{\hat{w}} = \argmax_{\mathbf{w}}{p(\mathbf{w} \mid \mathbf{X})} \propto \argmax_{\mathbf{w}}{p(\mathbf{w})p(\mathbf{X} \mid \mathbf{w})},
\end{equation}

\noindent where $p(\mathbf{w})$ defines the probability of each word sequence and is provided by a \emph{language model} that is independent of the type of biosignal used, while the likelihood $p(\mathbf{X} \mid \mathbf{w})$ is computed using \emph{acoustic and pronunciation lexicon models}. Of these, only the acoustic model depends on the specific biosignal, whereas the lexicon and language models are biosignal agnostic. In the following, we describe in more detail the problem of acoustic modelling for silent-speech \gls{asr}. We refer the interested reader to \cite{Rabiner93,Huang2001,Yu2014} for an introduction to \gls{asr} technology.

The term $p(\mathbf{X} \mid \mathbf{w})$ in \eqref{eq:asr} is what is known as the acoustic model in the \gls{asr} literature. It provides the likelihood of observing the sequence of feature vectors $\mathbf{X}$ under the assumption of the word sequence $\mathbf{w}$. In state-of-the-art \gls{asr} systems, each word $w$ is decomposed into a sequence of smaller subword units, such as phones or triphones, in order to reduce the data requirements when estimating the probabilities for systems with thousands of words. It also enables multiple pronunciations for each word. For this purpose, a dictionary is constructed with the phonetic transcription of each word supported by the system. Acoustic modelling, thus, is performed to estimate the probabilities of the observations for each subword unit. Typically, each subword unit is modelled with a \gls{hmm} \cite{Rabiner93} containing a fixed number of hidden states (e.g., 3 or 5 states), with each state corresponding to a stationary segment of the unit. Traditionally, each state emission distribution is modelled with a \gls{gmm} \cite{Deng1991}, although \glspl{dnn} were recently shown to achieve state-of-the-art performance in this task \cite{Hinton2012}.

Under the above assumptions, the computation of $p(\mathbf{X} \mid \mathbf{w})$ in \eqref{eq:asr} is carried out by marginalising over all \gls{hmm} state sequences corresponding to $\mathbf{w}$ as follows,
\begin{equation}
\label{eq:acoustic_model}
    p(\mathbf{X} \mid \mathbf{w}) = \sum_{\mathbf{Q}} p(\mathbf{X} \mid \mathbf{Q}) p(\mathbf{Q} \mid \mathbf{w}),
\end{equation}

\noindent where $\mathbf{Q} = (q_1, \ldots, q_T)$ is an \gls{hmm} state sequence, $p(\mathbf{X} \mid \mathbf{Q}) \approx \prod_{t=1}^T p(\bx_t|q_t)$ and $p(\mathbf{Q} \mid \mathbf{w}) \approx \prod_{t=1}^T p(q_t|q_t-1) $, assuming that the state sequence is a first-order Markov process (dependence on $\mathbf{w}$ has been excluded, for convenience).

Basically, biosignal-based \gls{asr} can be undertaken by replacing the front-end signal processing with techniques tailored to each specific biosignal, while the back-end acoustic modelling remains unchanged. This approach has been taken in several works, such as continuous phone-based \gls{hmm} recognition using \gls{semg} signals \cite{Jou2006TowardsCS} and isolated word recognition using image features for lipreading \cite{Chiou1997}. However, subword units in silent \gls{asr} tend to be harder to differentiate from other units than in audio-based \gls{asr}. For instance, in visual speech recognition, phones with a similar visual appearance but different acoustic characteristics are hard to distinguish by their visual characteristics alone. To address this issue, phones with similar articulatory properties are grouped to increase system robustness. For instance, phones with a similar visual appearance are grouped into \textit{viseme} units, or by considering articulatory gestures \cite{Cappelletta2011}. The main problem with this approach is the ambiguity that may be caused by visemes, which has to be resolved by language models. For \gls{emg}-based speech recognition, a data-driven approach called \textit{bundle phonetic features} was proposed in \cite{Schultz2010}. In general, biosignal-based speech recognition has been addressed via syllables \cite{Lopez-Larraz2010} and by using both context-dependent and context-independent phones \cite{Herff2015, Hueber2009}.

In addition, researchers have considered multimodal \gls{asr}, where multiple sources of information (e.g., brain signals, \gls{emg} onset, sound and muscle contraction) are combined to improve the recognition of spontaneous speech and increase robustness to noise \cite{King2007SpeechPK}. However, these sources are not synchronous \cite{Bregler1994EigenlipsFR} due to the multi-step nature of speech motor control and the complex relation between articulatory gestures and speech sounds \cite{Guenther2006}. Research is ongoing to resolve this issue.

Finally, the most likely word sequence $\mathbf{\hat{w}}$ given the sequence of feature vectors $\mathbf{X}$ is determined by searching all possible state sequences. An efficient way to solve this problem is to use the Viterbi algorithm \cite{Viterbi1967}, though several more efficient alternatives have been proposed, in which the breadth-first search of the Viterbi algorithm is replaced by a depth-first search \cite{Ney92}.


\subsection{Direct speech synthesis}
\label{ssec:ds}
Direct synthesis techniques are used to model the relationship between speech-related biosignals and the acoustic speech waveform. In its most common form, this relationship is conveniently represented as a mapping $\bff: \mathcal{X} \rightarrow \mathcal{Y}$ between the space $\mathcal{X} \in \mathbb{R}^{D_x}$ of feature vectors extracted from the biosignals and the space $\mathcal{Y} \in \mathbb{R}^{D_y} $ of acoustic feature vectors as follows:
\begin{equation}
    \by_t = \bff(\bx_t) + \beps_t,
    \label{equ:ds}
\end{equation}

\noindent where $\bx_t$ and $\by_t$ are, respectively, the source and target feature vectors at time $t$ computed from the silent speech and acoustic signals (more details about the computation of the source vectors is provided in Section \ref{sec:sensing_techniques}), and $\beps_t$ is a zero-mean independent and identically distributed (i.i.d.) error term.

Modelling the mapping function $\bff(\cdot)$ in \eqref{equ:ds} presents some challenging problems. This function is known to be non-linear \cite{Atal78,Neiberg2008}. Moreover, for some types of biosignals, this mapping is non-unique \cite{Qin2007,Neiberg2008,Ananthakrishnan2012}, that is, the same acoustic features may be associated with multiple realisations of the biosignal. For instance, ventriloquists are able to produce almost the same acoustics with multiple vocal tract configurations. Another reason for this non-uniqueness is that the sensing techniques frequently have a limited spatial or temporal resolution and, as a result, the speech production process is not properly captured and some information is lost.

Direct synthesis techniques can be classified as model-based or data-driven. With model-based techniques, it is assumed that the mapping in \eqref{equ:ds} can be described analytically using a closed-form mathematical expression. In general, however, this assumption only holds for certain articulator motion capture techniques, as those described in Section \ref{sec:articulatory_movement}. For these techniques, the mapping can be seen as a two-stage process: (i) vocal tract shape estimation and (ii) speech synthesis. Firstly, a low-dimensional representation of the vocal tract shape is derived from the captured articulatory data. For instance, in \cite{Toutios2011,Toutios2012,Toutios2013}, the control parameters of Maeda's articulatory model\footnote{Maeda's model is a 2D geometrical model of the vocal tract shape described using seven mid-sagittal control parameters: jaw opening, tongue dorsum position, tongue dorsum shape, tongue apex position, lip opening, lip protrusion and larynx height.} \cite{Maeda1990} were derived from the 3D positions of the lips, incisors, tongue and velum captured with \gls{ema} (see Section \ref{sssec:EPMA} for an in-depth description of \gls{ema}). Secondly, model-based  techniques generate the corresponding speech signal by simulating the airflow through the vocal tract model, a technique known as articulatory speech synthesis \cite{Fant1970,Rubin1981,Taylor2009}. Commonly, a digital filter representing the vocal tract transfer function is computed and, following the source-filter model of speech production \cite{Fant1970,Flanagan72,Rabiner78}, the acoustic waveform is finally synthesised by convolving the vocal-tract filter impulse response with the glottal excitation signal, which is normally approximated as white Gaussian noise for unvoiced sounds or an impulse train for voiced sounds. More advanced articulatory synthesis techniques have also been proposed, using realistic 3D vocal tract geometries in conjunction with numerical acoustic modelling techniques, such as the \gls{fem} \cite{Thomas1986,Oliveira2003,Arnela2016,Arnela2019} or the \gls{dwm} \cite{Murphy2007,Mullen2007,Speed2013,Gully2018}.

Although articulatory synthesis is the most natural and obvious way to synthesise speech, physical simulation of the human vocal tract presents some challenging problems. Firstly, the vocal tract model must be as accurate as possible in order to generate high-quality speech acoustics. On the other hand, the model should be simple enough to be implemented on a digital computer and have reasonable computational requirements. Unfortunately, these two conditions are often in conflict. For instance, although they are capable of generating high-quality speech, the computational load of the above-mentioned 3D \gls{fem} models of the vocal tract is prohibitive. Thus, computational times of 70-80 hours are reported in \cite{Arnela2016}, in which vowels of 20 ms were synthesised with the 3D \gls{fem} model, while in a more recent work \cite{Arnela2019} the same authors report an average time of six hours when simulating diphthongs with a duration of 0.2 s (with different computer specifications).

Because of these issues, the most successful direct speech synthesis techniques achieved so far in terms of speech quality are data-driven, in which the mapping in \eqref{equ:ds} is described as a multivariate regression problem. This mapping is usually modelled as a parametric function $\bff(\bx;\theta)$, where $\theta$ are the function parameters. In the \emph{training stage}, the parameters are estimated using a dataset with pairs of source and target vectors $\mathcal{D}=\lbrace (\bx_1, \by_1),\ldots, (\bx_N, \by_N)\rbrace$ derived from time-synchronous recordings of speech and biosignals obtained while the subject's voice is still intact or, at least, not severely impaired. In this case, the target vectors represent a compact acoustic parametrisation of the acoustic signal, such as \glspl{mfcc} \cite{Fukada1992} or \glspl{lsf} \cite{Mcloughlin2008}. Once the $\theta$ parameters have been estimated, the mapping function is deployed to restore the subject's voice by predicting the acoustic feature vectors from the biosignal. The final acoustic signal is then synthesised from the sequence of predicted acoustic feature vectors, using a high-quality vocoder (e.g., STRAIGHT \cite{Kawahara99}, WORLD \cite{Morise2016} or neural vocoders \cite{Tamamori2017,Ai2018}).


Various supervised machine learning techniques have been investigated to model the mapping function in \eqref{equ:ds}. Non-parametric machine learning techniques \cite[Ch.\ 18, p.\ 737]{Russell2010} such as shared Gaussian process dynamic models \cite{Gonzalez2015}, support vector regression \cite{Toutios2005} and a concatenative unit-selection approach \cite{Zahner2014,Janke2017,Herff2019}, have all been applied to this task, but by far the most successful techniques are those based on parametric methods. One such method is that of Gaussian mixture regression, where the joint probability function of source and target vectors is modelled by a \gls{gmm}, as follows:
\begin{equation}
    p(\bz) = \sum_{k=1}^K \pi^k \mathcal{N}(\bz;\bmu^k,\bSigma^k),
\end{equation}
\noindent where $\bz=(\bx^\top, \by^\top)^\top$ denotes the concatenation of the source and target feature vectors, $k=1,\ldots,K$ is the mixture component index, $\pi^k=P(k)$ denotes the prior probability of the $k$-th component, and $\mathcal{N}(\cdot)$ is a Gaussian distribution with mean vector $\bmu^k$ and covariance matrix $\bSigma^k$. In the training stage, the \gls{em} algorithm \cite{Dempster77} is used to optimise the \gls{gmm} parameters.

In the conversion stage, the acoustic features are predicted from the source features by computing the mean of the posterior distribution $p(\by|\bx_t)$. This value can be computed analytically as a linear combination of the posterior mean vectors of each Gaussian component, as described in \cite{Toda2007,Gonzalez2016}. This mapping algorithm has been used extensively for direct synthesis with different sensing technologies, such as \gls{pma} \cite{Gonzalez2016,Gonzalez2017}, \gls{ema} \cite{Toda2008}, \gls{emg} \cite{AToth2009,Janke2012,Janke2017} and non-audible murmur (NAM) \cite{Toda2005b,Toda2009,Toda2012}. Unfortunately, this algorithm presents the well-known issue that the trajectories of the estimated speech parameters contain perceivable discontinuities due to the frame-by-frame conversion process \cite{Toda2007}. To overcome this shortcoming \cite{Tokuda2000,Toda2007} proposed a trajectory-based conversion algorithm taking into account the statistics of the static and dynamic speech feature vectors. In particular, the joint distribution $p(\bx,\by,\bm{\Delta}\by)$ is modelled using a \gls{gmm}, where $\bm{\Delta}\by_t = \by_t - \by_{t-1}$ are the dynamic speech features (delta features) at frame $t$. In the conversion stage, the most likely sequence of static speech feature vectors is determined by solving the following optimisation problem:
\begin{equation}
    \hat{\bm Y}= \argmax_{\bm{Y}} p(\bm{W} \bm{Y}| \bm{X}),
    \label{equ:mlpg}
\end{equation}
\noindent where $\bm{X}=(\bx_1^\top, \bx_2^\top, \ldots, \bx_T^\top)^\top$ is the $(D_x T)$-dimensional sequence of source feature vectors, $\bm{Y}=(\by_1^\top, \by_2^\top, \ldots, \by_T^\top)^\top$ is the $(D_y T)$-dimensional sequence of static speech feature vectors to be determined, and $\bm{W}$ is a $(2 D_y T)$-by-$(D_y T)$ matrix representing the relationship between static and dynamic feature vectors such as $\overline{\bm Y} = \bm{W} \bm{Y}$, where $\overline{\bm Y}=(\by_1^\top,\bm{\Delta}\by_1^\top,\by_2^\top,\bm{\Delta}\by_2^\top,\ldots,\by_T^\top,\bm{\Delta}\by_T^\top)^\top$ is the $(2 D_y T)$-dimensional sequence of static and dynamic speech feature vectors. To solve the optimisation problem in \eqref{equ:mlpg}, an iterative EM-based algorithm was proposed in \cite{Toda2007}. This algorithm, known as the \gls{mlpg} algorithm, produces better acoustics than the conventional \gls{gmm} mapping described above because speech dynamics are also taken into account.

Apart from \glspl{gmm}, \glspl{hmm} have also been used for articulatory-to-acoustic conversion in the context of a multimodal \gls{ssi}, comprising video and ultrasound, with very promising results \cite{Hueber2011,Hueber2012,Hueber2016}. Another popular modelling technique is that of \glspl{dnn} \cite{Schmidhuber2015,Lecun2015}. Several works have reported evidence that \glspl{dnn} outperform mapping approaches like \glspl{gmm} and \glspl{hmm} in terms of conversion quality for \gls{emg} \cite{LDiener2015,JankeThesis,Janke2017}, \gls{pma} \cite{Gonzalez2017,Gonzalez2017b}, and in the related field of statistical voice conversion \cite{Desai2009,Mohammadi2014,Chen2014}, thanks to their powerful discriminative capabilities. For direct speech synthesis, various neural network architectures have been investigated, including feed-forward neural networks \cite{LDiener2015,Bocquelet2016,Gonzalez2017}, \glspl{cnn} \cite{diener2018session,Kimura2019,Juanpere2019} and (one of the most successful approaches) \glspl{rnn} \cite{Liu2016,Gonzalez2017b,Cao2018,Juanpere2019}. In \cite{Gonzalez2017} a comparison of different \gls{rnn} models for \gls{pma}-to-acoustic mapping was presented.

\subsection{Comparison of the two SSI approaches}
\label{ssec:ssi-comparison}
Each \gls{ssi} approach has its advantages and disadvantages. Silent speech-to-text has the advantage that speech might be more accurately predicted from the biosignals, thanks to the language and pronunciation lexicon models used in \gls{asr} systems. These models impose strong constraints during speech decoding and may help recover some speech features, such as voicing or manner of articulation, which are not well captured by current sensing techniques \cite{Hueber2010,Wand2011,Gonzalez2014,Gonzalez2017b}. However, the use of these models also means that this approach is unable to recognise words that were not considered during training, such as words in a foreign language. The direct speech synthesis approach, in contrast, is not limited to a specific vocabulary and is language-independent. A second limitation of the silent speech-to-text approach is that the paralinguistic features of speech (e.g., speaker identity or mood), which are important for human communication, are lost after \gls{asr}, but could be recovered by direct synthesis techniques. Yet another problem of silent speech-to-text is that, in practice, it is difficult to record enough silent speech data to train a large vocabulary \gls{asr} system\footnote{State-of-the-art \gls{dnn}-based \gls{asr} systems require hundreds of hours of carefully annotated speech data for training \cite{Hinton2012,Sainath2015,Chiu2018}.}, while direct synthesis systems require less training material (usually just a few hours of training data) because modelling the biosignal-to-speech mapping is arguably easier than training a full-fledged speech recogniser.

Nevertheless, the greatest disadvantage of the silent speech-to-text approach may be that it produces a disconnection between speech production and the corresponding auditory feedback, due to the long delay introduced by the \gls{asr} and \gls{tts} systems. In consequence, this approach lacks the real-time capabilities (i.e., low latency) that a \gls{ssi} system for natural human speech communication would require. In this regard, previous studies have estimated the maximum latency acceptable for an ideal \gls{ssi} system. In oral communication, 100 to 300 ms of propagation delay causes slight hesitation on a partner’s response and beyond 300 ms causes users to begin to back off to avoid interruption \cite{Na2002}. Studies of delayed auditory feedback, in which subjects receive delayed feedback of their voice, found disruptive effects on speech production with feedback delays starting at 50 ms, while delays of 200 ms produced maximal disruption \cite{Yates63,Stuart2002,Stuart2015}. Altogether, these results suggest an ideal latency of 50 ms for a \gls{ssi}, though latency values of up to 100 ms may still be acceptable. These low values can only be achieved through direct speech synthesis. In this sense, real-time \gls{ssi} systems have been developed for \gls{semg} \cite{LDiener2016An,LDiener2018cinves}, \gls{pma} \cite{Gonzalez2018} and \gls{ema} \cite{Bocquelet2016}. There is also the possibility that real-time auditory feedback might enable the brain to assimilate the \gls{ssi} as if it were the person's own voice, thus enabling the user to adapt her/his own speaking patterns to produce better acoustics. In this regard, previous \gls{bci} studies \cite{Lebedev2006,DiPino2009,Lebedev2014} have provided evidence of brain plasticity, enabling the gradual assimilation of assistive devices by the areas in the brain associated with motor control. 

\section{Sensing techniques}
\label{sec:sensing_techniques}
As illustrated in Fig. \ref{fig:biosignals}, the first step of any \gls{ssi} involves the acquisition of some kind of biosignal, different from the acoustic wave. These biosignals are the result of different activities (or processes) taking place in the human body during speech production, which can range from the movements of the articulators to neural activity in the brain. Thus, the production of the speech signal requires the movement of the different speech articulators (lips, tongue, palate, etc.) to shape the vocal tract, as well as the glottis and lungs. Muscles are responsible for these movements while the brain ultimately initiates, controls and coordinates them.

To monitor the speech production process, sensing techniques are used to acquire different types of biosignals related to this process. These biosignals can be recorded at the origin of the speech production, via sensing techniques for brain activity, or at the destination, by monitoring the resulting muscle activity. Alternatively, we can focus on the effects of muscle and brain activity and simply measure the movements of the articulators. In this section, we describe the sensor technology currently available and review previous \gls{ssi} research on each of the approaches proposed. 

\subsection{Brain activity}
Obtaining biosignals at the origin of speech production has the advantage that a wider range of speech disorders and pathologies can thus be addressed. Brain activity sensing techniques can potentially assist not only persons with voice disorders but also those with dysarthria or apraxia, or even some cases of aphasia. On the other hand, the internal processes of the brain that are involved in speech production are imperfectly understood, and recording brain activity at a high spatiotemporal resolution is still problematic, at best.

\subsubsection{Neuroanatomy of speech production}
The neuroanatomy of language production and comprehension has been a topic of intense investigation for more than 130 years \cite{Hickok2007}. Historically, the brain's left \gls{stg} has been identified as an important area for these cognitive processes. Studies have shown that patients with lesions to this brain area present deficits in language production and comprehension \cite{Damasio80}, and that a complex cortical network extending through multiple areas of the brain is involved in these processes \cite{Price12}.

This cortical network has recently been modelled by a dual-stream model consisting of a ventral and a dorsal stream \cite{Hickok2007}. The ventral stream, which involves structures in the superior (i.e., \gls{stg}) and middle portions of the temporal lobe, is related to speech processing for comprehension, while the dorsal stream maps acoustic speech signals to the frontal lobe articulatory networks, which are responsible for speech production. This dorsal stream is strongly left-hemisphere dominant and involves structures in the posterior dorsal and the posterior frontal lobe, including Broca's area, or \gls{ifg}, which is critically involved in speech production \cite{Flinker15}.

In the posterior frontal lobe, the cortical control of articulation is mediated by the ventral half of the lateral sensorimotor cortex or \gls{vsmc} \cite{Brown09}. This structure presents neural projections to the motor cortex of the face and the vocal tract and, by electrical stimulation, generates a somatotopic organisation of the face and mouth \cite{Bouchard2013}. However, focal stimulation of \gls{vsmc} does not evoke speech sounds, presumably because the production of phonemes and syllables requires multiple articulator representations across the \gls{vsmc} network coordinated in a certain motor pattern \cite{Bouchard2013}. This is consistent with an established neurocomputational model of speech motor control \cite{Guenther2006}, in which intended speech sounds are represented in terms of speech formant frequency trajectories. Projections from \gls{vsmc} to the primary motor cortex would transform the intended formant frequencies into motor commands to the speech articulators. This would be carried out in the same way as the desired three-dimensional spatial positioning of a fingertip is transformed into the angles of the corresponding articulation (shoulder, elbow, wrist, etc.) \cite{Guenther2009}.

\begin{table*}[t!]
\caption{Summary of sensor technologies available to monitor brain activity.}
\centering
\begin{tabular}{lccccc}
\textbf{Approach}  & \textbf{Sensing technique}               & \textbf{Temporal resolution} & \textbf{Spatial resolution } & \textbf{Intrusiveness} &  \textbf{Portability}\\ 
\hline
\hline
\multirow{2}{*}{Haemodynamic}    & \textit{fMRI}              &  Poor ($\geq$ 1 s)           & Good (1.5 - 2 mm)            & Non-intrusive           & Not portable\\
& \textit{fNIRS}                                            & Poor ($\geq$ 1 s)           & Medium (1 - 2 cm)              & Non-intrusive          & Moderate \\ 
\hline
\multirow{4}{*}{Electrodynamic} &  \textit{MEG}               & Good ($\geq$ 1 ms)       & Good (1 - 2 mm)              & Non-intrusive            & Not portable\\
& \textit{EEG}                & Good ($\geq$ 1 ms)       & Poor ($\approx$ 10 cm$^2$)                & Non-intrusive  & Moderate         \\
& \textit{ECoG}              & Good ($\geq$ 1 ms)       & Excellent (0.5 - 1 mm)        & Intrusive               & High  \\
&                                 \textit{LFP}               & Excellent ($\geq$ 0.1 ms)       & Excellent (5 -100 $\mu$m)        & Very intrusive & High        
\end{tabular}
\label{tab:bci_sensors}
\end{table*}

\subsubsection{Brain activity sensors}
\label{ssec:brain_sensors}
A range of sensors have been developed to capture neural activity during cognitive tasks. As shown in Table \ref{tab:bci_sensors}, these sensors essentially follow two main approaches to measure brain activity. In the first, the sensors measure the haemodynamic response (i.e., changes in blood oxygenation due to neuron activation) at certain locations of the brain. In the second, the sensors measure the electrodynamics in the brain, that is, the electrical currents and fields caused by activations during cognitive tasks.

Neuron activation requires the ions to be actively transferred across the neuronal cell membranes. The energy needed for this task is obtained through oxygen metabolism, which increases substantially during functional activation. By means of blood perfusion through the capillaries, oxygen is sent to the active neurons while the decrease in tissue oxygenation is counteracted by neurovascular coupling \cite{Ances2001}, a mechanism that regulates blood flow. This sequence of events produces changes in blood oxygenation, which are reflected in the balance of oxygenated (oxyHb) and deoxygenated (deoxyHb) haemoglobin, which is known as the \gls{hdr}. Various approaches can be employed to measure \gls{hdr}. For instance, oxyHb and deoxyHb haemoglobin have different magnetic properties that can be detected non-invasively by means of \gls{mri}. Doing so provides a three-dimensional high spatial resolution volume over the entire brain, which makes the \gls{fmri} the de-facto standard in neuroimaging \cite{Logothetis2008}. However, \gls{fmri} requires expensive and heavy equipment, which prevent this technique for being used as a wearable \gls{aac} device. \Gls{fnirs} is another non-invasive, moderately-portable technique that detects changes in \gls{hdr}, exploiting the fact that differences in absorptivity between oxyHb and deoxyHb and the transparency of biological tissue can be detected by means of infrared light emissions in the 700–1000 nm range \cite{Megan2014}. However, due to the limited penetration of infrared light into cerebral tissue, \gls{fnirs} imaging has a depth sensitivity of only about 0.75 mm below the brain surface \cite{Megan2014}. 

Methods based on the haemodynamic response are non-invasive and provide an excellent spatial resolution. Their main weakness is the temporal resolution, inherent to the neurovascular coupling, which is coarse and lagged. Thus, after the triggering event, the response lags for at least 1-2 s, peaks for 4-8 s and then decays for several seconds until homeostasis is restored \cite{Megan2014}. However, through trial repetition in which \gls{hdr} responses are combined, the temporal resolution can be improved to 1-2 s. In consequence, very few studies have considered their possible use for speech recognition and have achieved very limited success \cite{Schultz2017}.

Neuronal activity also provokes electric currents that can be measured in the extracellular medium. The synaptic transmembrane current, in the form of a spike, is the major contributor to this extracellular signal, although other sources can substantially shape it \cite{Buzsaki2012}. The superimposition of these currents within a volume of brain tissue generates electrical potentials and fields that can be recorded by electrodes. Such recordings have a time resolution of less than a millisecond, can be modelled reliably and are well understood \cite{Pesaran2018}. When these potentials are recorded using non-invasive electrodes from the scalp they are known as \gls{eeg}; or as \gls{ecog} when they are recorded from the cerebral cortex using invasive subdural grid electrodes \cite{Rabbani2019}; or as \glspl{lfp} when the measurement is obtained at deeper locations by inserting electrodes or probes \cite{Buzsaki2012}. Alternatively, currents generated by the neurons can be measured non-invasively outside the skull as ultra-weak magnetic fields. This technique, known as \gls{meg}, provides a relatively high spatiotemporal resolution ($\sim$1 ms, and 2–3 mm), as magnetic signals are much less dependent on the conductivity of the extracellular space than \gls{eeg} \cite{Hamalainen93}. Unfortunately, \gls{meg} requires expensive superconducting quantum interference devices operating in appropriate magnetically shielded rooms. Thus, \gls{meg} has been mostly used in neuroimaging and studies, although recently some attempts have been done in using this technique for silent speech recognition \cite{Dash2020}.

As a non-invasive technique, \gls{eeg} is the longest-standing and most widely used method for neural activity research \cite{Niedermeyer2005}. However, signals at \gls{eeg} electrodes are severely spatiotemporally smoothed and show little discernible relationship with the firing patterns of the contributing neurons. This is due to the large number of neurons involved in the recording and to the distorting and low-pass filter properties of the soft and hard tissues that the signal must penetrate before reaching the electrodes. Myoelectrical and environmental artefacts, as well as the subject's own movements, also distort the \gls{eeg} signal. For these reasons, \gls{eeg} is mainly used for \gls{aac}, which relies on time-locked averages or broad features of the neural firing signals, such as the P300 event-related potential, steady-state evoked or slow cortical potentials, and sensorimotor rhythms \cite{Brumberg2018}.

Invasive techniques, such as \gls{ecog} and \glspl{lfp}, despite the evident risks they pose, seem a more promising alternative for chronic implantation as the basis of a neural speech prosthesis \cite{Rabbani2019}. One such device is a probe designed at the University of Michigan \cite{Wise1970}, which is constructed on a silicon wafer using a photolithography process to pattern the interconnects and recording sites. This method allows electrode-tip diameters as small as 2 $\mu$m to be created, and facilitates controlled interelectrode spacing of 10 to 20 $\mu$m or greater. In addition, a semiflexible ribbon cable allows the probes to be suspended in the brain after they are inserted through the open dura. A different approach is taken with the probe array designed at the University of Utah \cite{Maynard1997}, which is fabricated from a solid block of silicon. Photolithography and thermomigration are used to define the recording sites and a micromachining process is then applied to remove all but a thin layer of silicon. During this process, eleven 1.5 mm-deep cuts are made along one axis; the wafer is then rotated through 90 degrees, and another eleven cuts are made. This results in a 10 $\times$10 array of needles with lengths ranging from 1.0 to 1.5 mm on a 4 $\times$4 mm square, which allows a large number of recordings to be obtained in a compact volume of the cortex. 

Another limitation of invasive techniques is that, as the electrode is inserted, some neurons are ripped while others are sliced. Moreover, blood vessels are damaged, provoking microhaemorrhages, initiating a signalling cascade and giving rise to the formation of a tight cellular sheath around the electrode after 6 to 12 weeks \cite{Schwartz2004}. This sheath eventually increases the impedance of the electrode, as the amount of exposed surface is compromised, preventing it from registering electrical activity. Despite this problem, some \gls{lfp} sensors, designed for longevity and signal stability, have been implanted successfully. One such is the neurotrophic electrode tested in \cite{Bartels2008}. This electrode uses a glass tip with a diameter of 50 to 100 microns which induces neurites to grow through it (three or four months after implantation). A disadvantage of this method is the fact that the number of wires in the probe is severely limited (to a maximum of four). On the other hand, the recordings last for the lifetime of the implant.

\subsubsection{SSIs based on brain activity signals} 
\glspl{bci} have been used for more than two decades to restore communication in severely paralysed individuals. Typical applications consist of a display presenting keyboard letters or pictograms that the user selects by forcing changes in their own electrophysiological activity. \glspl{eeg} signals, in their multiple variants (slow cortical potentials, P300 signal, sensorimotor rhythms, etc.) are commonly used to this end \cite{Brumberg2010,Lebedev2006,Wolpaw2002,Brumberg2018}. Unfortunately, these systems are very slow (one word or fewer per minute) and cognitively demanding, making conversational speech impractical. They are also difficult to master and require accurate sight. Conversely, \glspl{ssi} present a more practical and natural means of restoring speech communication abilities. Although much remains to be done, recent brain-sensing devices are paving the way for the introduction of these interfaces.

Despite the low spatial resolution of \gls{eeg}, some attempts have been made to synthesise speech from these signals. Thus, in \cite{Brumberg16}, continuous modulation of the sensorimotor rhythm is decoded into two-dimensional feature vectors with the first two formant frequencies, enabling real-time speech synthesis and feedback to the user. However, this approach relies on the activation of large areas of the cortex by imaging the movements of several limbs, and not by evoking speech production. In particular, participants in this study were instructed to imagining moving their right/left hand and feet when presented with vowel stimuli. In another study \cite{Krishna2020b}, speech synthesis from \gls{eeg} signals recorded in parallel while the participants either read aloud sentences or listen to pre-recorded utterances were investigated, achieving similar results in both conditions. Alternatively, the silent speech-to-text approach has also been investigated to decode speech from \gls{eeg} signals, but has encountered  the same limitations of low spatial resolution. Thee first attempts in this field were made by \cite{Suppes1997}, who proposed a word recogniser with a small vocabulary. Later attempts focused on phoneme \cite{Dasalla2009} and syllable \cite{DZmura2009} recognition but with a very limited dataset (3 phonemes and 6 syllables, respectively).

In recent years, significant advances have been achieved, following the use of invasive recording devices with better spatiotemporal resolution. In \cite{Lotte2015}, an \gls{hmm} classifier was used to model the mapping between continuous spoken phones and their \gls{ecog} representation. Following a similar approach, a `brain-to-text' system was presented in \cite{Herff2015} to decode speech from \gls{ecog} signals, which was able to achieve word and phone error rates below 25\% and 50\%, respectively. More recently, in \cite{Makin2020}, \emph{seq2seq} models were used to decode speech from \gls{ecog} signals, achieving WERs as low as 3\%.  In \cite{Herff2016}, a pilot study showed that \gls{ecog} recordings from temporal areas can be used to synthesise speech in real time. Data were collected from a patient fitted with bilateral temporal depth electrodes and three subdural strips placed on the cortex of the left temporal lobe while sentences were read aloud. Broadband gamma activity feature vectors extracted from the \gls{ecog} signals were mapped onto a log-power spectra representation of the speech signal. A simple sparse linear model was used to model this mapping. Although experimental results showed that the synthesised waveforms were unintelligible, broad aspects of the spectrogram were reconstructed and a promising correlation between the true and reconstructed speech feature vectors was observed. 

In \cite{Guenther2009}, a wireless \gls{bci} for real-time speech synthesis was proposed and tested for vowel production. A permanent neurotrophic electrode \cite{Bartels2008} was implanted in a peak activity region (revealed after a pre-surgery \gls{fmri} scan) on the precentral gyrus of a 26-year-old male volunteer who suffered from locked-in syndrome. To avoid wires passing through the skin and to minimise the risk of infection, \gls{lfp} signals were wirelessly transmitted across the scalp. These were later amplified and converted into frequency modulated radio signals. To decode speech, neural signals were processed by a Kalman filter to drive the first and second formant frequencies of a formant-based speech synthesiser (all other parameters were fixed) \cite{Guenther2009}. The whole decoding process was performed within 50 ms, enabling effective auditory feedback and accelerating the patient's learning process. 

More recently, in \cite{Anumanchipalli2019}, a direct speech synthesis system based on \gls{ecog} was proposed. \Gls{ecog} recordings were obtained by a high-density, $16 \times 16$ subdural electrode array placed over the left lateral surface of the brain. Although the grid placement was decided on purely clinical considerations (treatment for epilepsy), the study focused on five patients for whom the sensors covered brain areas involved in speech processing and production, namely the \gls{vsmc}, \gls{stg} and \gls{ifg} areas. Using time-synchronous \gls{ecog}-and-speech recordings from the patients while texts were read aloud, a two-stage deep learning-based system was trained to decode acoustic features from brain activity. In the first stage, articulatory kinematic features related to the position of the articulators were decoded by a neural network from \gls{ecog}. In the second stage, a set of acoustic speech features (\glspl{mfcc}, fundamental frequency (F0) and voicing and glottal excitation gains) were decoded by a second bidirectional \gls{rnn} from the articulatory features predicted in the first stage. Listening tests conducted over the resulting synthesised speech signals revealed that, given a closed vocabulary (25 and 50 words), listeners could readily identify and transcribe the speech signals. A similar approach is followed in \cite{Angrick2019}, where \gls{ecog} recordings from an $8 \times 8$ electrode grid placed over the ventral motor cortex, the premotor cortex and the \gls{ifg}, for a group of six patients, were transformed into speech features (logMel spectrogram) using a \gls{cnn} trained with parallel data involving time-synchronous \gls{ecog}-and-speech recordings. A second \gls{dnn} was then used to synthesise speech from these features. Experimental results showed that the intelligibility of the decoded signals (measured through objective metrics) was over 50\% in some cases. Moreover, in \cite{Herff2019}, it was shown that \gls{tts} decoding strategies can be applied to the same recordings, resulting in more intelligible speech and enabling real-time implementation of the synthesiser.

\subsection{Muscle activity}
As shown in Fig. \ref{fig:biosignals}, during speech production the muscles in the face and larynx are responsible for the movements that will eventually result in the production of the acoustic signal. As mentioned above, the brain controls the activation of these muscles by means of electrical signals transmitted through the motor neurons of the peripheral nervous system. These electrical signals cause muscles to contract and relax, thus producing the required articulatory movements and gestures. \Gls{emg} measures the electrical potentials generated by depolarisation of the external membrane of the muscle fibres in response to the stimulation of the muscles by the motor neurons \cite{Tapia2020}. The \gls{emg} signal resulting from the application of this technique is complex and dependent on the anatomical and physiological properties of the muscles \cite{Raez2006}.

\begin{figure}[t]
    \centering
    \includegraphics[width=\linewidth]{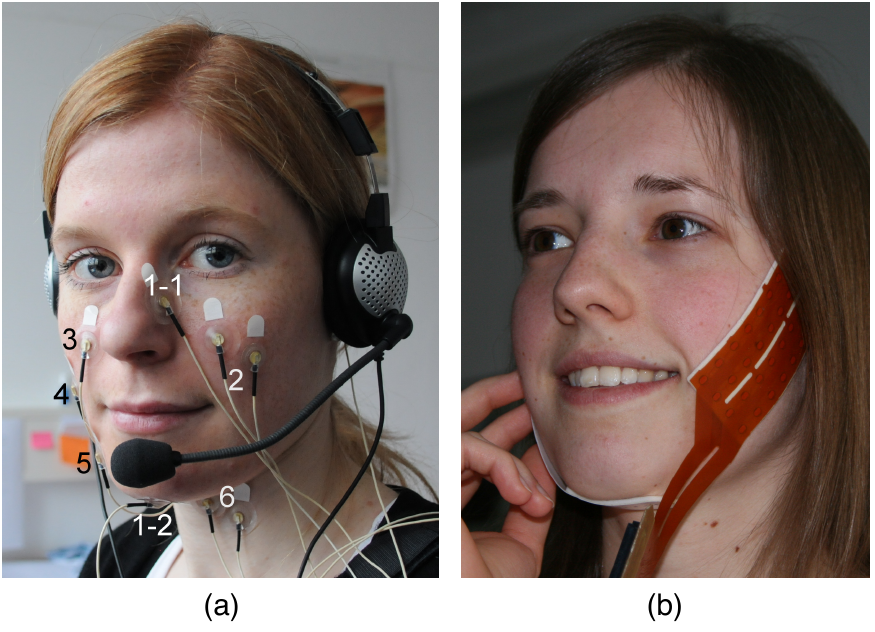}
    \caption{\gls{semg} sensor devices. (a) Single electrodes. (b) Arrays of electrodes. With the permission of \cite{LDiener2015direct}.}
    \label{fig:sEMG}
\end{figure}

 
Two types of electrodes can be used for \gls{emg} signal acquisition: invasive or non-invasive. Invasive methods involve intramuscular electrodes (i.e., needles) inserted through the skin into the muscle. These methods fundamentally measure localised action potentials, but this approach can be problematic when the aim is to measure the characteristics and behaviour of whole muscle signals, as is the case with \glspl{ssi}. In contrast, non-invasive methods employ superficial electrodes (i.e., \gls{semg}) directly attached to the skin, as shown in Fig. \ref{fig:sEMG}. In this case, the \gls{semg} signal is a composite of all the action potentials of the muscle fibres localised beneath the area covered by the sensor. Because of this property and its non-invasiveness, \gls{semg} is the preferred technology in most \gls{ssi} investigations. The characteristics of \gls{semg} signals are determined by the properties of the tissue separating the signal generating sources from the surface electrodes. In particular, biological tissue acts as a low pass filter affecting the frequency content of the signal and the distance at which it can no longer be detected.

\subsubsection{EMG and speech production}
In studies of speech production and related applications, \gls{emg} electrodes are attached to the subject's face, as illustrated in Fig. \ref{fig:sEMG}. Fig. \ref{fig:sEMG}a shows a single electrode setup \cite{MaierHein2005, YDeng2014} with electrodes connected to certain muscle areas, whereas Fig. \ref{fig:sEMG}b shows an electrode array setup \cite{Wand2013array,LDiener2015direct}. In the latter case, there are two electrode arrays, a large one placed on the cheek and a small one under the chin. The signals thus captured represent the potential differences between two adjacent electrodes. Once amplified, these signals are ready for further signal processing.


Since the speech signal is mainly produced by the activity of the tongue and the facial muscles, the \gls{emg} signal patterns resulting from measurements in these muscles provide a means of retrieving the speech signal \cite{Chan2001}. Moreover, this effect is maintained even when words are spoken inaudibly, i.e., the acoustic signal is not produced \cite{Jorgensen2003neu}. This represents an important advantage of \gls{emg}-based \gls{ssi} systems when it comes to providing an alternative means of communication for persons with voice disorders (such as laryngectomy patients) or some types of speech disorders (e.g., dysarthria). Another advantage is that \gls{emg} signals appear 60 ms before articulatory motion \cite{Netsell1975,Jou2006}, which is an important feature for real-time \gls{emg}-to-speech conversion with low latency.




Besides its application in \glspl{ssi} (see next section), \gls{emg} is being used in clinical rehabilitation (e.g., for the recovery of facial muscular activity in patients with motor speech disorders \cite{Draizar1984} and other articulatory disturbances \cite{Daniel1978}), assistance and as an input device \cite{Tapia2020}.
In particular, these previous studies have reported the benefits of \gls{emg} biofeedback in therapy aimed at increasing muscle activity of the oral articulators in dysarthric speakers with neurological conditions \cite{Netsell1973,Draizar1984,Nemec1984}. \gls{emg} is also a useful tool for speech production research \cite{Minifie1974,Blair1986}.


\subsubsection{SSIs based on EMG signals}
The first studies of \gls{emg} for speech recognition date back to the mid-1980s. These initial studies were conducted on very small vocabularies consisting of just a few words or commands \cite{sugie1985speech, morse1986research, morse1989use, morse1991speech}. Thanks to this very limited vocabulary, the recognition accuracy achieved in these works was high \cite{chan2003multi, Jorgensen2003neu}. Subsequently, \gls{emg}-based speech recognition of complete sentences was addressed in \cite{Jou2006} with an acceptable recognition rate (\textasciitilde 70\% in a single-speaker setup). To enable \gls{emg}-based speech recognition with large vocabularies, several subword units were investigated in \cite{Walliczek2006sub,Jou2006}, including a data-driven approach known as \emph{bundled phonetic features} \cite{Schultz2010,WandTesis}, which models the interdependences between phonetic features (voiced, alveolar, fricative, etc.) using a decision-tree clustering algorithm. More recently, hybrid \gls{emg}-\gls{dnn} systems for \gls{emg}-based \gls{asr} were investigated in \cite{MWand2016deep}. In \cite{Wang2020}, transfer learning was found to be beneficial for silent speech recognition from \gls{emg} signals by exploiting neural networks trained on a image classification task as powerful feature extraction models. More recently, in \cite{Wang2020b}, an empirical study was conducted to investigate the effect of the number of \gls{semg} channels in silent speech recognition.

Direct speech synthesis from \gls{emg} signals has also progressed considerably in recent years (see \cite{LDiener2015,LDiener2016An,Janke2017,LDiener2018cinves}), following advances in array \gls{semg} sensors and deep learning. As mentioned above, a particular advantage of \gls{emg} with respect to other techniques for articulator motion capture is that \gls{emg} signals can be sensed \textasciitilde 60 ms before the actual movements of the articulators. This rapidity facilitates the development of real-time direct synthesis systems with low latency \cite{LDiener2016An,LDiener2018cinves}, so that the delay between the articulatory gestures and the synthesised acoustic feedback is minimal. In \cite{LDiener2018cinves}, a comprehensive study was carried out in which the influence of various system parameters (\gls{dnn} size, amount of training data, frame shift, etc.) on the speech quality generated by a real-time direct synthesis system was analysed using objective quality metrics.

Although significant progress has been made, \gls{ssi} technology based on \gls{semg} still faces several issues, which are currently under intense investigation. One such issue is the strong dependence of the results on the training session. Although this effect can be reduced by using array \gls{semg} sensors (such that the relative position of the sensors is kept constant), there are still differences between data captured in different sessions \cite{Janke2017, diener2018session}. To address this issue, an unsupervised adaptation technique was proposed in \cite{MWand2014} allowing new data to be incorporated with each data recording session. More recently, in \cite{MWand2018}, a domain-adversarial training approach \cite{YGanin2015} was investigated to adapt the front-end of \gls{emg}-based speech recognition systems to the target session data in an unsupervised manner. Besides, discrepancies between audible and silent speech articulation are also known to influence these systems. In particular, in \cite{Wand2014} it was shown that \gls{asr} performance is severely degraded in mismatched conditions. Ultimately, this effect was attributed to differences in the spectral content of the \gls{semg} signals captured for different modes of speaking. To address this issue, a spectral mapping algorithm was proposed, aiming to transform \gls{semg} data obtained during silently mouthed speech, so that the transformed data would resemble data obtained during audible speech articulation. After applying this technique in combination with multi-style training, an improvement of 14.3\% in recognition rates was achieved, obtaining an average \gls{wer} of 34.7\% for silently mouthed speech and 16.8\% for audibly spoken speech. Another important issue that has yet to be resolved is that of speaker independence. To date, most studies have been carried out in a speaker-dependent fashion, meaning that the \gls{asr} (or direct synthesis) models are trained using only data from the end user. Enabling speaker independence would allow these models to be trained more robustly, requiring fewer data from the end user. This is further discussed in Section \ref{ssec:challenges_inter_intra_variability}.


\subsection{Articulator motion capture}
\label{sec:articulatory_movement}
Production of the acoustic speech signal requires the movement of different speech articulators. Therefore, monitoring the movement of these articulators is a straightforward approach enabling us to capture meaningful biosignals for speech characterisation. In this subsection, we describe different techniques for capturing articulatory movement, using kinematic sensors attached to the vocal tract or by means of imaging techniques to visualise these changes. As most of these techniques do not capture glottal activity, they are best suited to restore communication capabilities for persons with voice disorders, such as laryngectomised patients.

\subsubsection{Magnetic articulography}
\label{sssec:EPMA}
The techniques described in this section employ magnetic tracers attached to the articulators and sense their movement by measuring the changes in the magnetic field generated (or sensed) by these magnets. There are two variants of this technique, \gls{ema} and \gls{pma}, which differ according to where the generation and sensing of the magnetic field take place. A comparative study of these variants can be found in \cite{Cao19}.

The idea of \gls{ema} \cite{Schonle87,Hummel2006} is to attach receiver coils to the main articulators of the vocal tract. These coils are connected by wires attached to external equipment that monitors articulatory activity. Transmitter coils placed near the user's head generate alternating magnetic fields, making it possible to track the spatial position of the coupled receiver coils. The advantages of this technique are its high temporal resolution for modelling the articulatory dynamics and the minimal feature pre-processing required (the captured data directly provides the 3D Cartesian coordinates of the receiver coils and, additionally, their velocity and acceleration). The major drawback is the need for external non-portable transmitters and wired connections, which limits its use to laboratory experiments.

\Gls{ema} was used in \cite{Heracleous2011} for automatic phoneme recognition without additional audio information. A Carstens AG100 device simultaneously tracked the vertical and horizontal coordinates in the mid-sagittal plane of six receiver coils located at different points of the oro-facial articulators. The \gls{ema} parameters were recorded at a sampling frequency of 500 Hz. The articulatory parameters were fused using a multi-stream \gls{hmm} decision fusion. This study was conducted using a French corpus of vowel-consonant-vowel (VCV) sequences and additional short and long sentences. Finally, the system was evaluated using different combinations of articulatory parameters and compared with the use of the standalone audio signal or a combination of both audio and \gls{ema} data. The recognition results obtained were found to be competitive. \glspl{dnn} were recently investigated in the context of \gls{ema}-based \gls{asr}. In \cite{Kim2017}, bidirectional \glspl{rnn} were used to capture long-range temporal dependencies in the articulatory movements. Moreover, physiological and data-driven normalisation techniques were considered for speaker-independent silent speech recognition. A silent speech \gls{ema} dataset was recorded from twelve healthy and two laryngectomised English speakers. This approach provided state-of-the-art performance in comparison with other \gls{asr} models.

\Gls{ema} data have also been employed for speech synthesis. In \cite{Toda2008}, the \gls{gmm}-based conversion algorithms described in Section \ref{ssec:ds} were applied to \gls{ema}-to-speech and speech-to-\gls{ema} (i.e., articulatory inversion) tasks using the MOCHA database \cite{MOCHA}. Experimental results demonstrated the superiority of the \gls{mlpg}-based mapping algorithm for both tasks compared to conventional \gls{mmse}-based mapping. In \cite{Aryal2016}, an alternative modelling approach based on a tapped-delay input line \gls{dnn} was explored, seeking to improve \gls{ema}-to-speech mapping accuracy by capturing more context information in the articulatory trajectory. Subjective evaluation showed a strong preference for the \gls{dnn} approach, in comparison with previous \gls{gmm}-based approaches. An extension to bidirectional \glspl{rnn} was proposed in \cite{Liu2018} and an augmented input representation was also investigated to deal with the known limitations in data acquisition technology. One problem with the above approaches is that they do not consider the differences in articulatory movements between neutral and whispered speech. To address this issue, a transformation function was investigated in \cite{Meenakshi2018} to reconstruct neutral speech articulatory trajectories from whispered ones. The results showed that an affine transformation can satisfactorily approximate the relation between the two speaking modes. More recently, in \cite{Stone2020b}, pitch prediction (i.e., prediction of the speech voicing and fundamental frequency) from \gls{ema} data captured by six coils placed on the upper lip, the lower lip, the lower incisor, the tongue tip, the tongue body, and the tongue dorsum was investigated, achieving surprisingly good results despite \gls{ema} not capturing any information about the vibrations of the vocal folds.

Besides their use in data-driven articulatory synthesis, \gls{ema} data have also been employed in standard \gls{tts} systems as a means of improving the naturalness of synthesised speech and enhancing system flexibility. Thus, in \cite{Ling2009}, several approaches were investigated to integrate \gls{ema} articulatory features into these systems. The accuracy and naturalness of the predicted acoustic parameters were improved with the integration of these articulatory features. Furthermore, this integration enabled a degree of control over the acoustic parameter generation process. 

\begin{figure}[t]
    \centering
    \includegraphics[width=\linewidth]{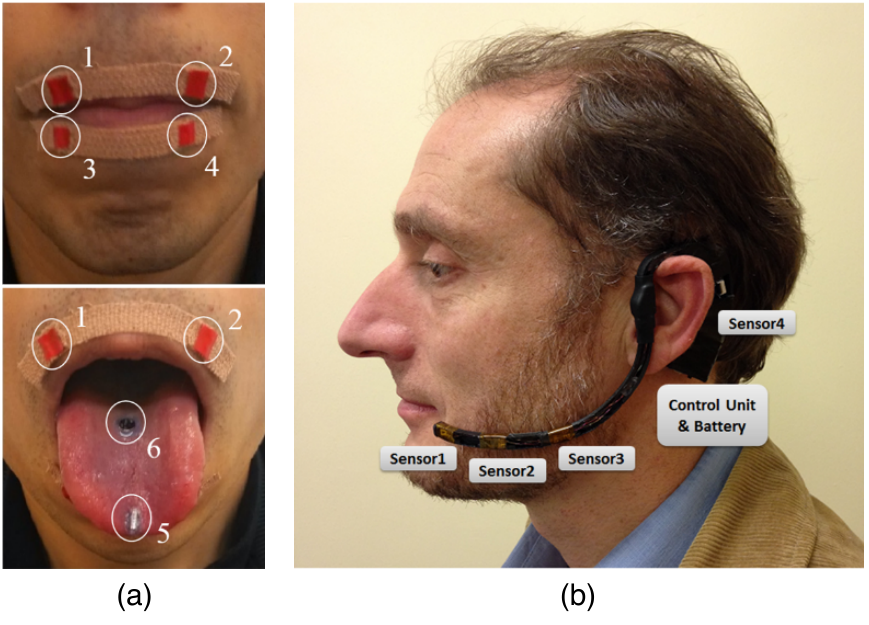}
    \caption{\gls{pma} sensor device. (a) Placement of the magnetic pellets on the tongue and lips. (b) Wearable sensor headset to measure articulator movements. With the permission of \cite{Gonzalez2017}.}
    \label{fig:pma}
\end{figure}

The second variant of the technique for capturing articulator movement using magnetic tracers is \gls{pma} \cite{Fagan2008}. In this technique, several small permanent magnets are attached to a set of points in the vocal articulators. The sum of the magnetic fields generated by these magnets is measured by magnetic sensors placed outside the mouth, as shown in Fig. \ref{fig:pma}. Among the advantages of \gls{pma} in comparison to \gls{ema}, it does not require wired connections and the sensors are easy to place. This makes the technique more comfortable for the user and facilitates portability. Nevertheless, the data thus acquired are a composite of all the magnetic fields generated by the magnets, and so their relation with the spatial position of the magnetic tracers is less explicit in \gls{pma}. In consequence, additional pre-processing is needed \cite{Sebkhi2019}. 

The first attempt to develop a speech recognition system using \gls{pma} was carried out in \cite{Fagan2008}. This paper proposed a simple system for recognising isolated words based on the \gls{dtw} algorithm. The \gls{pma} setup consisted of seven small magnets temporarily attached to the user's lips and tongue, together with a sensing system of six dual-axis magnetic sensors incorporated into a pair of glasses. A total of twelve outputs from these sensors were captured at a sample rate of 4kHz. The user was asked to repeat a set of nine words and thirteen phones. The patterns were recognised with an accuracy of 97\% for words and 94\% for phonemes. A similar approach was followed in \cite{Gilbert2010}, achieving recognition rates of over 90\% for a 57-word vocabulary. In \cite{Hofe2013}, an \gls{hmm}-based speech recognition system from \gls{pma} data was described. The system was evaluated on recognition tasks both for isolated words and for connected digits. Feed-forward \glspl{dnn} for \gls{pma}-based speech recognition were recently evaluated in \cite{Kim2018}, who reported an average phoneme error rate of 37.3\% and a \gls{wer} of 32.1\%.

Direct speech synthesis from \gls{pma} data was first evaluated in \cite{Hofe2011}, in which speech formants were estimated from articulatory data using a simple linear model. In \cite{Gonzalez2016}, a more complex model was investigated to model the mapping between the articulatory and acoustic feature spaces. A mixture of factor analysers (MFA) was used, approximating the mapping function in a piece-wise linear fashion. During the conversion phase, the acoustic parameters were estimated from the \gls{pma} feature vectors by using the \gls{mmse} or the \gls{mle} estimation procedures introduced in Section \ref{ssec:ds}. Recent studies, \cite{Gilbert2017, Gonzalez2017}, have evaluated more complex models, such as \glspl{gmm}, \glspl{dnn} and \glspl{rnn}, for the \gls{pma}-to-speech task. Speech signals generated by these models were (on average) \textasciitilde 75\% intelligible (as measured by human listeners), but for some participants intelligibility scores reached \textasciitilde 92\%.

\subsubsection{Palatography}
Electropalatography (EPG) \cite{Hardcastle1989} is a technique for recording the timing and location of tongue contacts with electrodes placed in a pseudo-palate inside the mouth during speech production. The pattern of palatal contacts provides information about the articulation of different phones. Optopalatography (OPG) \cite{Wrench1998} is a similar technique which uses optical distance sensors, making it possible to record the tongue position and lip movements without requiring explicit contact with the palate. 

Most studies of these techniques have been conducted in the fields of speech therapy and phonetic research. For instance, in \cite{Toutios2008}, a data-driven approach was used to map the speech signal onto EPG contact information by means of \gls{pca} and support vector machine (SVM). In \cite{Birkholz2011}, information about the articulation of vowels and consonants was obtained by means of a new sensing technique known as electro-optical stomatography (EOS), which combines the advantages of EPG and OPG. This technique was later evaluated for vowel recognition in \cite{Birkholz2012}, using EPG patterns and tongue contours as features and a \gls{dnn}-based classifier. This research was extended in \cite{Stone2016,Stone2020} to enable the recognition of German command words. In \cite{Mumtaz2014}, EOS data were used to reconstruct the tongue contour using a multiple linear regression model. A problem with OPG is that an error is introduced if the tongue is not oriented perpendicular to the axes of the optical sensors. To overcome this error, Stone et al. \cite{Stone2017} proposed a model of light propagation for arbitrary source-reflector-detector setups which considered the complex reflective properties of the tongue surface due to sub-surface scattering.

\subsubsection{Imaging techniques}
The use of video and imaging techniques is a simple, direct way to obtain information about the movement of the external articulators, such as the lips and jaw. Moreover, a variety of audio-visual data corpora \cite{Cooke2006,Chung17} have been developed in recent years and are freely available for research purposes. This has boosted the development of audio-visual speech systems \cite{Katsaggelos2015,Abdelaziz2018,Afouras2018} for tasks such as recognition, synthesis and enhancement. In addition, recent studies have explored the use of visual-only information \cite{Petridis2017,Petridis2018}, which provides only partial information about the speech production process, and \gls{rtmri} of the vocal tract \cite{Leeuwen2019}, which enables the acquisition of 2D magnetic resonance images at an appropriate temporal resolution (typically, 5-50 frames per second) to examine vocal tract dynamics during continuous speech production. The combination of video and ultrasound imaging to capture the movement of the intraoral articulators (the tongue, mainly) has achieved promising results in the context of \gls{ssi} research \cite{Hueber2010,Schultz2017}. Radar sensors have also been investigated in the context of \gls{ssi} research as a promising alternative to capture articulator motion data \cite{Lee2019,Lee2020}. In this subsection, we focus on imaging techniques that are suitable for speech recognition/synthesis in the absence of the acoustic speech signal.

Ultrasound imaging was first used for speech synthesis in \cite{Denby2004}. An Acoustic Imaging Performa 30 Hz ultrasound machine, placed under the user's chin, was used for tracking tongue movement. This provided a partial view of its surface in the mid-sagittal plane. A multilayer perceptron (MLP) was then used to map the captured tongue contours to a set of acoustic parameters driving a speech vocoder (see Section \ref{ssec:ds}). In \cite{Denby2006}, a similar approach was followed but additional lip profile information extracted from video was employed, and the resulting articulatory features were mapped to \gls{lsf} speech features. A new approach, called \textit{Eigentongue} decomposition, was proposed in \cite{Hueber2007}, improving upon previous approaches with respect to tongue contour extraction from ultrasound images.

In \cite{Hueber2010}, a segmental vocoder driven by ultrasound and optical images of the tongue and lips was proposed. Visual features extracted from the tongue and lips were used to train context-independent continuous \glspl{hmm} for each phonetic class. During the recognition stage, phonetic targets were identified from these visual features and an audio-visual unit dictionary was used for corpus-based speech synthesis. The speech waveform was generated by concatenating acoustic segments with a prosodic pattern using a harmonic plus noise model \cite{Stylianou1995}. The system was evaluated using an hour of continuous audio-visual speech obtained from two speakers, achieving 60\% correct phoneme recognition. Later, in \cite{Hueber2016}, a direct synthesis technique was employed, using an \gls{hmm}-based regression technique with full-covariance multivariate Gaussians and dynamic features.

Recent advances in deep learning for image processing have paved the way for the application of these techniques in \gls{ssi} research. For instance, in \cite{Xu2017}, a \gls{cnn} was used to classify tongue gestures, obtained by ultrasound imaging during speech production, into their corresponding phonetic classes. \Glspl{cnn} for visual speech recognition were also evaluated in \cite{Tatulli2017}. A multimodal \gls{cnn} was used to jointly process video and ultrasound images in order to extract visual features for an \gls{hmm}-\gls{gmm} acoustic model. This model achieved a recognition accuracy of 80.4\% when tested over the database developed in \cite{Hueber2016}, which validated it for visual speech recognition. Deep autoencoders were used in \cite{Ji2018,Gosztolya2019} to extract features from ultrasound images, achieving significant gains in both silent \gls{asr} and direct synthesis. In \cite{Toth2018}, multitask learning of speech recognition and synthesis parameters was evaluated in the context of an ultrasound-based \gls{ssi} system designed to enhance the performance of individual tasks. The proposed method used a \gls{dnn}-based mapping which was trained to simultaneously optimise two loss functions: an \gls{asr} loss, aiming at recognising phonetic units (corresponding to the states of an \gls{hmm}-\gls{dnn} recogniser) from the input articulatory features; and a speech synthesis loss, which predicted a set of acoustic parameters from the input features. Using the proposed scheme, a relative error rate reduction of about 7\% was reported both for speech recognition and for speech synthesis. Finally, convolutional recurrent neural networks were recently employed in \cite{Zhao2019,Xu2019} for tongue motion prediction and feature extraction from ultrasound videos.

\section{Current challenges and future research directions}
\label{sec:challenges}
\Glspl{ssi} have advanced considerably in recent years \cite{Denby2017}. The studies reviewed in the previous sections show that it is possible to decode speech, even in real-time in some cases, for a wide range of speech-related biosignals. Nevertheless, this technology is not yet mature enough for useful purposes outside laboratory settings. In particular, most \glspl{ssi} have been validated only for healthy individuals, and their viability as a clinical tool to assist speech-impaired persons has yet to be determined. Furthermore, to the best of our knowledge, none of the studies we reference report a means by which intelligible speech can be consistently generated, in the context of a large vocabulary and/or conversational speech. In addition, there remains the problem of training data-driven direct synthesis techniques when no acoustic data are available.

In order to move from the laboratory to real-world environments, several challenges need to be addressed in future research, including:
\begin{itemize}
    \item The development of improved sensing techniques capable of recording biosignals with sufficient spatiotemporal resolution to enable the decoding of speech with minimal delay. The biosignals captured should provide sufficient information about the speech production process to enable the intended words to be decoded with high accuracy and, possibly also, to provide other important information, such as prosody or paralinguistics. As a clinical device, these sensing techniques must be safe for long-term implantation or, for non-invasive devices, sufficiently robust to be worn every day.
    
    \item The development of robust, accurate algorithms for decoding speech from the biosignals, by automatic speech recognition or direct speech synthesis. These algorithms should be flexible enough to resolve diverse clinical scenarios, ranging from individuals with mild communication disorders from whom training material (biosignals and possibly speech recordings) can be easily obtained, to patients who are completely paralysed and from whom little or no data will be available. In these circumstances, \emph{a priori}, data recording could be an exhausting process. Furthermore, direct synthesis algorithms should be capable of synthesising speech with high intelligibility and naturalness, ideally resembling the user's own voice.
    
    \item To date, most studies in this field have evaluated \gls{ssi} systems \emph{offline}, using pre-recorded data. However, future investigations will need to evaluate these systems under more challenging \emph{online} scenarios, such as the \emph{user-in-the-loop} setup, in which users receive real-time acoustic feedback on their actions, so that both the users and the \gls{ssi} can exploit this feedback to improve their performance. Furthermore, previous investigations in the related field of \glspl{bci} \cite{Lebedev2006} have shown that brain plasticity enables prosthetic devices in such a user-in-the-loop scenario to be assimilated as if they were part of the subject's own body. However, this assimilation has yet to be investigated for the case of \gls{ssi}-based speech prosthetic devices.
    
    \item As clinical devices intended for use as an alternative communication method for persons with communication deficits, it is still necessary to determine the role that \glspl{ssi} will play in \gls{slt} practice.
\end{itemize}

In the following subsections, we discuss some key points regarding the first three technical challenges outlined above. The fourth, the clinical aspect of the question, lies beyond the scope of this paper.

\subsection{Improved sensing techniques}
Most of the sensing techniques described in Section \ref{sec:sensing_techniques} have only been validated in laboratory settings under controlled scenarios. Hence, certain issues need to be addressed before final products can be made available to the general public. First, while many techniques are designed to allow some portability and to be generally non-invasive, some problems remain. The equipment is not discreet and/or comfortable enough to be used as a wearable in real-world practice \cite{Cheah2016,Liu2020} and may be insufficiently robust against sensor misalignment \cite{Wand2011session,Gonzalez2016b}. Second, the linguistic information captured by these devices is often limited. For example, \gls{semg} has difficulty in capturing tongue motions, while \gls{ema}/\gls{pma} cannot accurately model the phones articulated at the back of the vocal tract due to practical problems that may arise in locating sensors in this area (such as the gag reflex and the danger of the user swallowing the sensors) \cite{Gonzalez2014,Gonzalez2017b}. These problems might be overcome by combining different types of sensors, each of which is focused on a different region of the vocal tract, thus enabling a broader spectrum of linguistic information to be obtained. Yet another issue is that of how to capture and model supra-segmental features (i.e., prosodic features), which play a key role in oral communication. Prosody is mainly conditioned by the airflow and the vibration of the vocal folds, which in the case of laryngectomised patients is not possible to recover. As a result, most direct synthesis techniques generating a voice from sensed articulatory movements can, at best, recover a monotonous voice with limited pitch variations \cite{Gonzalez2017,Diener2019,Csapo2019}. The use of complementary information capable of restoring prosodic features is thus an important area for future research.

Another key issue is the need to develop wireless sensors, thus eliminating cumbersome wired connections. Although some sensors with these communication capabilities can be found (e.g., wireless \gls{semg} sensors \cite{BTSBioengineering,DelsysInc,Shimmer}), they have yet to achieve acceptable levels of miniaturisation and robustness making them suitable for everyday use. Another practical design question is the need to reduce energy consumption and to increase sensor use time between battery charges. A problem related to that of energy consumption is the need to establish an efficient, low-power communication protocol. In this respect, various alternatives have been proposed, such as Bluetooth Low Energy \cite{Gomez12}, IEEE 802.15.6 \cite{Std802.15.6} and LoRa \cite{Vangelista15}.

While most of the above discussion is focused on non-invasive sensing techniques, invasive techniques such as those employed in \glspl{bci} pose even greater challenges. Apart from the issues mentioned above (portability, wireless communication, energy consumption, etc.), a long-awaited feature for \gls{bci} systems is the availability of biocompatible sensors capable of providing long-term recordings from multiple cortical and sub-cortical brain areas \cite{Lebedev2006}. In order to precisely monitor the complex electrical activity underlying speech-related cognitive processes, neural probes with enormous spatial density, multiplexed recording array integration and a minimal footprint are required. In recent years, several neural probes have been developed in projects addressing these issues, and the results obtained have surpassed conventional medical neurotechnologies by many orders of magnitude. For example, the NeuroGrid array is a recently developed \gls{ecog}-type array, fabricated on flexible parylene substrates using photolithographic methods, which covers a 10 mm x 9 mm area with 360 channels \cite{Khodagholy2014}. Alternatively, the Neuropixels probe \cite{Jun2017} features 960 recording electrodes on a single shank, measuring 10 mm by 70 $\mu$m, while the Neuroseeker probe \cite{Raducanu17} contains 1344 electrodes on a shank of 50 $\mu$m x 100 $\mu$m and 8 mm long, providing the greatest number of independent recording sites per probe to date. On the other hand, Neuralink \cite{Musk2019} offers 3072 recording channels enclosed in a package of less than 23 mm x 18.5 mm, which is able to control 96 independent probes with 192 electrodes each. In addition, a robot capable of inserting six of these probes per minute with micron precision and vasculature avoidance has been developed by the same company.

Stable, chronically-implanted probes can be achieved by minimising the immune system's reaction to the implant and by reducing the relative shear motion at the probe-tissue interface. In this respect, mesh electronic neural probes \cite{Hong2018} feature a bending stiffness comparable to that of brain tissue whilst offering sufficient macroporosity to enable neurite interpenetration, thus preventing the accumulation of pro-inflammatory signalling molecules. A radically different approach is adopted in \cite{Adewole2018}, where `living electrodes' based on tissue engineering are proposed. Although only a proof of concept has so far been offered, the authors of this paper have developed axonal tracts \emph{in vitro} which might be optically controlled \emph{in vivo}. To date, these novel electrode technologies \cite{Hong2019} have only been tested on animals, but eventually, the leap to humans will be made, making it possible to record even single-unit spiking activity from individual neurons.

\subsection{Training with non-parallel data}
The data-driven direct synthesis techniques described in Section \ref{ssec:ds} assume the availability of time-synchronised recordings of biosignal and speech from the patient. This scenario, however, does not cover the whole spectrum of clinical scenarios that might be found in real life. For instance, for a given individual there may exist enough pre-recorded speech data while no time-aligned biosignal recordings are available. On the opposite side of the spectrum, there may be paralysed individuals from whom no previous speech recordings are available, but who could benefit from \gls{ssi} technology to speak again.

To accommodate all of these scenarios, new training schemes must be developed. Voice banking \cite{Yamagishi2012,Veaux2013,Erro2015}, i.e., capturing and storing voices from voice donors so they can later be used to personalise \gls{tts} systems, will certainly play a major role in the development of these algorithms, providing the acoustic recordings necessary to train the mappings. Once appropriate speech recordings are available (provided by a voice donor or previously recorded by the user), biosignals covering the same phonetic content as the speech recordings must be captured from the end-user (asked to silently mouth during reproduction of the speech recordings). Mappings could then be trained to translate the biosignals into audio, as described in Section \ref{ssec:ds}. However, the direct application of standard supervised machine learning techniques might not be feasible if (as is highly likely) the recordings for the two modalities have different durations. In this case, deep multi-view time-warping algorithms \cite{Vu2012,Trigeorgis2017,Tung2019} could be applied to align the modalities and thus achieve effective mapping. As an alternative, sequence-to-sequence mapping techniques \cite{Sutskever2014,Vaswani2017} represent another interesting possibility to model the mapping in the absence of parallel recordings. However, in general, the application of sequence-to-sequence techniques would preclude real-time speech synthesis, as these techniques normally process the whole sequence at once. Yet another interesting alternative is to apply non-parallel adversarial neural architectures based on the popular CycleGAN technique \cite{Zhu2017}. These architectures, which have been successfully applied in the related field of voice conversion \cite{Kameoka2018,Fang2018}, make it possible to train mappings when neither parallel data nor time alignment procedures are available.

The need to obtain time-synchronous audio recordings from the user could be avoided by deploying model-based direct synthesis techniques rather than data-driven approaches. The former, as discussed in Section \ref{ssec:ds}, employ articulatory models to synthesise speech by simulating the airflow through a physical model of the vocal tract. These techniques are able to generate speech, provided that the shape of the vocal tract can be easily recovered from the biosignals. This, however, is only the case for certain types of articulator motion capture techniques (e.g., \gls{ema} \cite{AlBawab2009,Toutios2011,Toutios2012,Toutios2013}). For other types of sensing techniques, the application model-based direct synthesis techniques would imply, as a prerequisite, training a mapping from the biosignals to an intermediate representation with information about the articulator kinematics \cite{Anumanchipalli2019}.

\subsection{Online and incremental training algorithms}
Current \glspl{ssi} assume that the distribution of biosignal features will remain unaltered during both training and evaluation, making them unable to adapt to changes in patterns of use over time. However, these changes are likely to occur, either as a deterioration of the user's skills due to progressive illness or with improvement, as the user gradually masters the \gls{ssi} with practice. In order to cope with these changes, therefore, new training algorithms requiring little or no manual intervention and minimal labelled data are needed. Reinforcement learning algorithms \cite[Ch.\ 21]{Russell2010} \cite{Mnih2015} offer a promising alternative to supervised learning algorithms, enabling the control of \glspl{ssi} to be continuously optimised in response to user feedback. These algorithms have been applied in previous research into the online control of robotic prosthesis \cite{Pilarski2011} and in computational models imitating vocal learning in infants \cite{Murakami2015,Warlaumont2016}, with encouraging results.

Besides coping with changes in use patterns over time, algorithms must also be developed to enable incremental learning and control of \glspl{ssi}. It seems impractical (not to mention exhausting for the user) to suggest that all the data required to train a full-fledged \gls{ssi} device could be recorded in a single session. A more realistic setup would be to record enough training material to establish an initial system that is simple enough for good performance to be expected (e.g., a system only able to decode vowels or isolated words) but at the same time, one that is genuinely useful. With this, we seek to avoid user frustration and early abandonment of the \gls{ssi} because it does not meet the user's expectations and communication needs \cite{Christensen2012}. From this initial system, we should aim at creating a `virtuous circle' whereby practice improves the user's control over the \gls{ssi}, which in turn provides more data for \gls{ssi} training. When this \gls{ssi} reaches peak performance, the functionality of the system could be gradually expanded, e.g., to decode a richer vocabulary or to work with continuous speech.

\subsection{Robustness against inter- and intra-subject variability}
\label{ssec:challenges_inter_intra_variability}
Most of the \gls{ssi} systems are speaker and session-dependent, meaning that they are trained with data recorded for one particular subject in a given recording session. While this approach may be reasonable for systems with chronically implanted sensors, where the biosignal distribution is not expected to vary greatly between different uses of the system, inter-session variability will certainly affect systems based on wearable sensors. Inter-session variability refers to changes in the statistical distribution of the biosignal features as a result of sensor repositioning between system training and use of the \gls{ssi}, or changes in the physical properties of the human body due to disease or to external factors (such as the weather). Inter-session variability is known to have a detrimental effect on the performance of \gls{ssi} systems, both for silent approaches \gls{asr} \cite{Wand2011session,MWand2014} and for direct synthesis \cite{Gonzalez2016b,Gosztolya2020}. Various methods have been proposed to increase the robustness of \glspl{ssi} against this type of variability, including feature-level adaptation \cite{Gonzalez2016b,Bocquelet2016}, model adaptation \cite{MWand2014,Gosztolya2020}, multi-style training \cite{Wand2011session} or the use of a domain-adversarial loss function to integrate session independence into neural network training \cite{MWand2018}. Despite these advances, session independence for \glspl{ssi} using wearable sensors is far from being fully achieved.

Another type of intra-subject variability is that of speaking mode variations, which arise from differences in the cognitive and/or physical processes involved in different ways of speaking. These differences are reflected in the biosignals acquired, causing the distribution of signal features to differ significantly among speaking modes. In consequence, the performance of \glspl{ssi} is degraded when training is performed on data captured from a given speaking mode (e.g., audible speech articulation) and evaluation is conducted on a different one (e.g., silently mouthed speech). One of the simplest, yet most reliable ways to counteract these differences is by training the system with data from multiple speaking modes (i.e., multi-style training). For \gls{emg}-based speech recognition, another technique that has proved useful is to use a spectral mapping algorithm designed to reduce the mismatch between the \gls{semg} signals of audible and silent speech articulation \cite{Janke2012,Wand2014}. In the case of \glspl{ssi} based on brain activity signals, although audible speech production and inner (imagined) speech are very different cognitive processes, they share common neural mechanisms \cite{Aleman2004,Geva2011}. However, in an experiment conducted in \cite{Martin2014} in which models trained with \gls{ecog} data captured during audible articulation were used to reconstruct speech for the covert condition (i.e., inner speech condition), it was shown that reconstruction accuracy was significantly lower than for the matched condition. This performance reduction was attributed to differences between the two speaking modes.

Inter-subject variability refers to differences due to anatomical variability among subjects. These differences are reflected in the speech and in the biosignals acquired from different subjects. Hence, the direct application of a \gls{ssi} trained for one subject to another will result in many errors. Achieving speaker-independence, thus, is a major and long-standing goal in \gls{ssi} research, as this would allow new systems to be bootstrapped requiring only a small fraction of training data from the end-user. This issue is particularly relevant for persons with moderate to severe speech impairments, for whom data collection can be exhausting. While most of the \glspl{ssi} proposed to date largely rely on speaker-dependent models, some recent studies have investigated ways of reducing across-subject variability. In \cite{Wang2014}, the normalisation of \gls{ema} articulatory patterns across subjects was achieved through the application of Procrustes matching, which is a bidimensional regression technique for removing the translational, scaling and rotational effects of spatial data. Following this articulatory normalisation, speech recognition accuracy for a speaker-independent system trained with data from multiple subjects improved significantly, from 68.63\% to 95.90\%, becoming almost equivalent to a speaker-dependent system. Later, in \cite{Wang2015,Kim2017}, Procrustes matching was combined with speaker adaptive training (SAT) to further improve the results. The experimental results showed that, after applying Procrustes matching and SAT, recognition accuracy for a speaker-independent system improved significantly, in most cases outperforming a speaker-dependent baseline system. Another approach that has proved successful in improving speaker-independence in \gls{asr} systems is to use a domain-adversarial loss function during neural network training \cite{Wand2017}, which causes the network to learn a speaker-agnostic intermediate feature representation. Recently, in \cite{Dash2019}, both supervised and unsupervised adaptation strategies were investigated for enabling speaker-independency when decoding continuous phrases from MEG signals. The results indicate that the adaptation strategies improved significantly the recognition results.

Despite these advances, no significant achievements have been made towards achieving speaker-independence in direct synthesis techniques. Although some of the above-mentioned techniques could be applied to obtain an input feature representation independent of the subject, direct speech synthesis techniques would also benefit from adapting the audio output to make it resemble the user's own voice. This might be achieved by means of speaker adaptation algorithms originally developed for speech synthesis \cite{Yamagishi2007,Yamagishi2009,Wu2015}. The aim of these techniques is to adapt an average voice (or speaker-independent) model trained with speech from multiple speakers to sound like a target speaker using a small adaptation dataset from that speaker. To adapt such a model, various techniques have been developed, such as \gls{mllr} based speaker adaptation \cite{Leggetter1995} or, for \gls{dnn}-based parametric speech synthesis, augmenting the neural network inputs with speaker-specific features (e.g., i-vectors \cite{Saon2013}).

\subsection{Validation in a clinical population}
With few exceptions \cite{Meltzner2017sil,Kim2017}, research outcomes in \glspl{ssi} have been validated only for healthy users. Although clinical populations have participated in studies with implanted brain sensors, in most cases the participants have had the sensors implanted to treat other neurological conditions (e.g., refractory epilepsy). Furthermore, in these cases sensor implantation was driven by clinical needs, with little or no consideration for optimising sensor placement to cover the language-processing areas in the brain. 

In most cases, therefore, \glspl{ssi} have been validated only for healthy users. This is for several reasons: first, \gls{ssi} development is still in its infancy. Thus, many of the studies reviewed in this paper only seek to show that speech can be decoded from a particular biosignal for healthy individuals. Second, data recording is considerably harder for patients with health conditions than for healthy individuals. In particular, the synchronous recording of parallel audio-and-biosignal data may not be feasible for persons with moderate to severe speech impairments. As mentioned above, the non-availability of parallel training data greatly hampers the use of direct synthesis techniques for this population. Third, persons with speech impairments are likely to display more variability than those with no such impairments. Thus, biosignal variability as the impairment advances is another type of intra-subject variability which should be considered in practical \glspl{ssi}. However, designing systems that are robust to such variability is no easy task. Finally, difficulties may arise in recruiting patients for the studies and in addressing the ethical questions involved. 

\subsection{Evaluation in more realistic scenarios}
The vast majority of \glspl{ssi} thus far proposed have been validated using offline analyses with pre-recorded data. In these analyses, a pre-recorded data corpus is used both for system training and for evaluation. While the results of these offline analyses are useful for optimising various system parameters (such as system latency, output quality and system robustness), online analyses are needed in order to evaluate system performance in real-world scenarios. Online analyses assess the efficacy of the \gls{ssi} while it is in active use, possibly while the user is receiving real-time audio feedback. Ideally, the system should be tested in real-life scenarios, over a prolonged period (i.e., longitudinal analysis) and with an adequate number of users presenting a diversity of speech impairments at different stages of evolution. Regarding the first point, most offline analyses reported to date have been based on a pre-recorded list of words, commands or phonetically-rich sentences. While this type of vocabulary-oriented evaluation can provide insights into \gls{ssi} accuracy for decoding different phones, it does not reflect the fact that, in most cases, users will employ the \gls{ssi} to establish a goal-oriented dialogue (e.g., ordering food in a restaurant or asking for help) \cite{Moses2019,Li2020}. In these situations, other factors come into play, such as contextual information and visual clues (e.g., body language), which can help to resolve confusion in word meaning during the dialogue \cite{Gonzalez2018}.

\subsection{Availability of public datasets}
One of the factors that is slowing the development of \gls{ssi} technology is the lack of large datasets, which are required for developing speech tools and are very time-consuming to collect. Accordingly, most studies conducted in this field have used small datasets recorded by different research groups, using in-house biosignal recording devices. This diversity of approach has led to research fragmentation, making it difficult to compare the technical and algorithmic advances of different technologies. 

To our knowledge, the only large database available, with the required characteristics, is TORGO \cite{TORGO}, which contains about 23 hours of time-aligned acoustic and \gls{ema} articulatory signals obtained from eight dysarthric speakers and seven controls. Although alternative datasets exist, such as the MOCHA \gls{ema} articulatory database \cite{MOCHA}, the \gls{emg}-UKA parallel speech-and-\gls{emg} corpus \cite{EMGUKA} and the Wisconsin X-ray microbeam (XRMB) corpus \cite{XRMB}, they only contain data for healthy speakers. Certainly, data collected from healthy individuals can be used to develop technology for speech-impaired people. However, this type of data does not reflect the great variability that is present in speech-impaired subjects arising from two main sources: (i) inter-person variations according to the disorder and its severity; (ii) intra-person variability as the disease progresses. Furthermore, it is always a challenge to record a sufficient amount of data for persons with neurological and physical disabilities.

\section{Concluding remarks}
\label{sec:conclusions}
In this paper, we review recent attempts to decode speech from non-acoustic biosignals generated during speech production, ranging from capturing the movement of the speech articulators to recording brain activity. We present a comprehensive list of sensing technologies currently being considered for capturing biosignals, including invasive and non-invasive (wearable) techniques. From these biosignals, speech can be decoded by automatic speech recognition (ASR) or by direct speech synthesis. A potential advantage of the latter approach is that it may enable speech to be decoded in real time. This means that the direct synthesis approach might be able to restore a person's original voice, while the \gls{asr} approach, at best, would be like having an interpreter.

Although researchers have shown that it is indeed possible to decode speech from biosignals, the performance and robustness offered by present-day \glspl{ssi} remain insufficient for their large-scale use outside laboratory settings. We highlight several crucial factors that are still preventing the widespread implementation of this technology, including the need for better sensors, for new training algorithms, for non-parallel and zero-data scenarios, and for systems to be validated for use in clinical populations. When these challenges are overcome, \glspl{ssi} could become a real communication option for persons with severe communication deficits. We expect significant advances in all these directions in the years to come.

\section*{Acknowledgements}
This work was funded by Agencia Estatal de Investigación (AEI) under the grant PID2019-108040RB-C22/AEI/10.13039/501100011033. Jose A. Gonzalez-Lopez holds a Juan de la Cierva-Incorporation Fellowship from the Spanish Ministry of Science, Innovation and Universities (IJCI-2017-32926).

\Urlmuskip=0mu plus 1mu\relax
\bibliographystyle{ieeetran}
\bibliography{journal_abrv_short,paper}

\end{document}